\documentclass[prx,superscriptaddress,reprint]{revtex4-2}

\usepackage[titletoc,toc,title,page]{appendix}
\usepackage{graphicx} 
\usepackage{mathtools}
\usepackage{amsmath}
\usepackage{cases}
\usepackage{calc}
\usepackage[usenames,dvipsnames]{color}
\usepackage{mathrsfs} 
\usepackage[normalem]{ulem} 

\usepackage{nameref}
\usepackage[colorlinks=true]{hyperref}
\usepackage[nameinlink]{cleveref}
\crefname{appsec}{Appendix}{Appendices}
\crefname{box}{Box}{Box}
\hypersetup{
  colorlinks   = true, 
  urlcolor     = green!80!black, 
  linkcolor    = blue, 
  citecolor    = red!80!black, 
  breaklinks=true   
}
\usepackage{xcolor} 
\usepackage{physics} 
\usepackage{float} 
\usepackage{graphicx}
\usepackage[usenames,dvipsnames,table]{xcolor}
\usepackage{easyReview}
\usepackage{comment}
\usepackage{tikz}
\usetikzlibrary{calc,shapes.geometric}
\usepackage{amsmath,amssymb,amsthm,thmtools}
\usepackage{multirow,tabularx,booktabs}
\usepackage{makecell}
\usepackage[multiple]{footmisc}
\usepackage{bm}
\usepackage{bbm}
\usepackage{soul}
\setstcolor{red}

\usepackage{lipsum}

\begin{document}
\title{Revealing high-dimensional entanglement through symmetry}

\author{Jayden Webster}
\thanks{These authors contributed equally to this work.}
\affiliation{Quantum and Advanced Technologies Research Institute, Griffith University, Yuggera Country, Brisbane, QLD 4111, Australia}

\author{Florian Kanitschar\textcolor{blue}{$^*$}}
\email{florian.kanitschar@outlook.com}
\affiliation{Atominstitut Technische Universität Wien, Stadionallee 2 1020, Vienna, Austria}
\affiliation{AIT- Austrian Institute of Technology, Center for Digital Safety and Security, Giefinggasse 4, 1210 Wien, Austria}

\author{Emanuele Polino}
\email{e.polino@griffith.edu.au}

\author{Simon J. U. White}
\affiliation{Quantum and Advanced Technologies Research Institute, Griffith University, Yuggera Country, Brisbane, QLD 4111, Australia}

\author{\\Sven Rogge}
\affiliation{School of Physics, University of New South Wales, Sydney, NSW 2052, Australia}

\author{Farzad Ghafari}
\affiliation{Quantum and Advanced Technologies Research Institute, Griffith University, Yuggera Country, Brisbane, QLD 4111, Australia}

\author{Marcus Huber}
\email{marcus.huber@tuwien.ac.at}
\affiliation{Atominstitut Technische Universität Wien, Stadionallee 2 1020, Vienna, Austria}
\affiliation{Institute for Quantum Optics and Quantum Information (IQOQI), Austrian Academy of Sciences, Boltzmanngasse 3, 1090 Vienna, Austria}

\author{Nora Tischler}
\affiliation{Quantum and Advanced Technologies Research Institute, Griffith University, Yuggera Country, Brisbane, QLD 4111, Australia}

\begin{abstract}
Photons encoded in discrete time bins can be routinely prepared in temporal superposition states, enabling high-dimensional entanglement and enhanced quantum communication rates. However, characterizing this high-dimensional entanglement presents significant challenges, namely due to the involved measurement complexity or reliance on restrictive assumptions that compromise the generality of traditional approaches. Here, we develop and experimentally demonstrate a simple linear-optical scheme based on particle-exchange symmetry that allows us to probe high-dimensional entanglement in time-bin-encoded states. Combining Hong-Ou-Mandel interference with suitable transformations, our method not only certifies entanglement but also lower-bounds its dimensionality using only two dichotomic symmetry-based measurements. This bound is obtained through a new rigorous theoretical analysis and can be further improved by weak, physically motivated assumptions. The scheme remains effective at any timescale, even far below the temporal detector resolution used. Our work provides a powerful state-characterization tool and demonstrates that we can prove high-dimensional temporal entanglement on timescales inaccessible to the setup.
\end{abstract}
\maketitle

\section{Introduction}
Entanglement \cite{horodecki2009quantum} stands as one of the most striking phenomena of quantum mechanics, giving rise to correlations beyond any classical description \cite{brunner2014bell,wiseman2017causarum}. In addition to its foundational significance, entanglement is key for achieving quantum advantages in communication \cite{Bennett_Brassard_1984, Ekert_1991, xu2020secure,azuma2023quantum, bergmayr2023harness, kanitschar2024practical, Kanitschar_2025}, metrology \cite{polino2020photonic,barbieri2022optical}, and computation \cite{nielsen2010quantum,preskill2023quantum}. When extended to high (beyond two) dimensions, entanglement acquires even greater potential: enhanced information capacity, loss tolerance, and noise resilience all open new foundational and technological directions \cite{erhard2020advances,wang2020qudits,zhu2021high,zahidy2024practical}. In parallel, time has always been a central yet elusive concept in science \cite{rovelli2019order}, and its role in quantum theory remains an open question \cite{muga2007time}. Despite these conceptual challenges, encoding in time provides a powerful way to overcome the fragility of entanglement, enabling the creation of robust, high-capacity quantum states \cite{brendel1999pulsed,richart2012experimental,zhang2014unconditional,reimer2019high} for a wide range of quantum information tasks \cite{singh2025photonic,xavier2025energy}. The most effective strategy is to harness the full spatio-temporal structure of single photons, thereby maximizing the information per transmitted particle. Conveniently, spontaneous parametric down-conversion (SPDC) naturally provides entanglement across these degrees of freedom, with the temporal mode already demonstrating outstanding performance in both fiber~\cite{tittel1998violation,marcikic2003long,marcikic2004distribution,cuevas2013long,sun2016quantum,yu2025quantum,montaut2025progress} and free space~\cite{brendel1999pulsed,tittel2000quantum,zhong2015photon,islam2017provably,cozzolino2019high,fitzke2022scalable,valivarthi2020teleportation,wen2022realizing} links. 

However, characterizing entanglement in high dimensions is a significant challenge due to the number of measurements required for a complete state reconstruction. This issue is particularly acute for time-bin qudits, where quantum information is encoded in discretized windows of the photon's arrival time. In the temporal domain, measurements typically rely on unbalanced Mach-Zehnder interferometers, as in the Franson scheme \cite{franson1989bell,takesue2009implementation,ikuta2022scalable}. These interferometers impose several limitations: Coherence can be probed only at discrete time separations defined by the interferometer arms, and stabilizing multiple interferometers simultaneously is technically demanding. Even if the interferometer is appropriately sized, the minimum temporal distance between time bins is usually constrained by the detector's time resolution. Other approaches based on nonlinear optics, such as pulse shaping, sum-frequency generation \cite{pe2005temporal,kuzucu2008time,donohue2013coherent,maclean2018direct} or hybrid interferometric-nonlinear schemes \cite{bouchard2022quantum,bouchard2023measuring}, offer alternative means, as do other measurement strategies that combine both time and frequency \cite{friis2019entanglement}. However, all of these approaches remain complex and tend to introduce strong assumptions about both the probed state and detectors. While introducing assumptions about the state can reduce the complexity of the required measurements, such assumptions can lead to inaccurate predictions when applied incorrectly, compromising the generality of the approach. So far, no efficient and experimentally robust method has been demonstrated for certifying high-dimensional time-bin entanglement without such experimental or theoretical constraints.

In this work, we address these challenges by developing and experimentally realizing an alternative approach based on a fundamental concept in physics: particle exchange symmetry \cite{sakurai2020modern}. It is known that Hong-Ou-Mandel (HOM) interference \cite{hong1987measurement,fabre2022hong,bouchard2020two} can be used to witness entanglement \cite{benatti2020entanglement} by observing anti-symmetry in photonic states \cite{Freq1}. This connection between entanglement and anti-symmetry in photonic states has been explored across multiple platforms, from frequency-domain implementations in bulk and integrated optics \cite{Freq1,Freq2,barbieri2017hong,graffitti2020direct,chen2021temporal, eckstein2008broadband, kaneda2019direct, ramelow2009discrete, maltese2020generation, merkouche2022heralding, chen2022telecom, chen2020verification}, to transverse spatial modes \cite{TransverseSM}, transverse momentum modes \cite{gao2022manipulating} and orbital angular momentum \cite{OAM1,OAM2,OAM3} entangled states. However, accessing the dimensionality of entanglement through standard HOM anti-symmetric projections is impossible.

Here, we show that symmetry enables us to go far beyond mere entanglement certification: With only two dichotomic, symmetry-based measurements, we provide a direct route to quantifying high-dimensional entanglement. We derive a semidefinite program (SDP) combined with a novel Schmidt-number witness constrained by those measurement outcomes and transform it into its dual. This allows us to certify high-dimensional entanglement and provide a rigorous lower bound on the entanglement dimension with minimal assumptions about the underlying quantum state. Aided by additional physically motivated symmetries of the unknown quantum state, valid in non-adversarial settings, we show how the lower bound on the entanglement dimensionality can be improved. We implement this scheme in a compact photonic experiment that exploits the high-dimensional entanglement naturally generated in SPDC, with measurements performed through simple linear-optical transformations followed by HOM interference. From this experiment, we certify entanglement dimensions of 12 and 3 for the physically justified and adversarial scenarios, respectively.
Our approach explicitly distinguishes the treatment of the incoming state and the measurement apparatus, since the objective is to infer properties of an unknown incoming quantum state using a trusted measurement device, with the linear-optical transformations forming an integral part of the measurement apparatus. This scenario is particularly relevant in settings where the measurement apparatus is under the experimenter's control, whereas the photon source may belong to a third party.

The proposed scheme enables the manipulation of the exchange symmetry of high-dimensional time-bin states, thereby facilitating the rigorous certification of their entanglement dimension. Furthermore, it relies on only two dichotomic measurements and allows time-bin separations independent of the temporal detector resolution, even, in principle, below any currently achievable detection timescale. These features establish the method as a versatile and powerful tool for studying high-dimensional quantum systems and their application to quantum technologies.

\section{Symmetry-based measurements on high-dimensional photonic states}
To begin, we discretize the time domain into intervals $\{t_1, t_2, \dots, t_d\}$, where $d$ is the dimension of each subsystem, and refer to the corresponding basis states by kets $\ket{1}, \ket{2}, ..., \ket{d}$. Discretization provides a convenient representation of the state, but also has other implications: it inherently limits the maximum amount of entanglement contained within any given subspace of the discretized Hilbert space, and determines how experimental imperfections, such as detector timing jitter and noise, affect the observed correlations. As a result, even when extremely high-dimensional entangled states are produced by a physical process such as SPDC, access to this entanglement depends on the choice of discretization. In any chosen time-bin basis discretization, an arbitrary state $\rho$ can be expressed as:

    \begin{equation}\label{mixedstate}
    \rho = 
    \sum_{i,j,k,l=1}^{d} 
    \rho_{ij,kl} \, |i\rangle_1 \langle j| \otimes |k\rangle_2 \langle l|,
    \end{equation}
    
where $\rho_{ij,kl} \in \mathbb{C}$ are the complex matrix elements of the density operator in the computational basis, and the subscripts 1 and 2 refer to the first and second photon, respectively. The goal of this work is to probe this time-bin state and lower-bound its entanglement dimension. More specifically, in what follows, we quantify the entanglement content of this state via two dichotomic measurements, inspired by fundamental symmetry transformations. To achieve this, our scheme can be split into three steps:

    \begin{enumerate}
        \item Manipulate the symmetry of the state through suitable transformations.
        \item Perform symmetry projections using HOM interference.
        \item Extract the entanglement dimension from these projections through an SDP approach. 
    \end{enumerate}

To illustrate this first step, consider an example using the following maximally entangled pure state:

\begin{equation}\label{targetstate}
	\ket{\Phi^+}= \frac{1}{\sqrt{d}} \sum^{d-1}_{i=0} \ket{i}_1\ket{i}_2.
\end{equation}

This state is the one naturally generated by SPDC pumped by a continuous-wave (CW) laser under ideal conditions, truncated to a limited number of time-bins. To match the experimental data reasonably well, this truncation should be shorter than the coherence time of the laser. Each of the measurements on our state first involve a transformation, $U_\pm$, which acts on both photons: For the first photon, $U_\pm$ transforms each single time-bin into a superposition of two bins, $\ket{i}_1 \rightarrow \frac{1}{\sqrt{2}}\big(\ket{i}_1 \pm \ket{i+2}_1\big)$, while the second photon is delayed by a single bin $\ket{i}_2 \rightarrow \ket{i+1}_2$. These two transformations can be described by the following relations, when acting on the computational time-bin basis $\{\ket{i}_1,\,\ket{j}_2\}$ where the indices $i,j \in \{ 0,...,d-1\}$:

\begin{equation}\label{symtransforms}
\begin{split}
    U_\pm\ket{i}_1\ket{j}_2 = \frac{\ket{i}_1\pm\ket{i+2}_1}{\sqrt{2}} \ket{j+1}_2 \;.
\end{split}
\end{equation}

Applying the operations of Eq.~\eqref{symtransforms} on the state from Eq.~\eqref{targetstate}, we obtain:

\begin{equation}\label{finalentstate}
\begin{split}
	&U_\pm\ket{\Phi^+}= \frac{1}{\sqrt{d}}\sum^{d-1}_{i=0} \frac{\ket{i}_1\pm \ket{i+2}_1}{\sqrt{2}}  \;\ket{i+1}_2 \; =\\
    &\frac{1}{\sqrt{2d}}\Big[\ket{0}_1\ket{1}_2 +\Big(\pm\ket{2}_1\ket{1}_2 +\ket{1}_1\ket{2}_2+\cdot \cdot \cdot\\
    & \pm\ket{d}_1\ket{d-1}_2 +\ket{d-1}_1\ket{d}_2\Big) \pm\ket{d+1}_1\ket{d}_2\Big]\;.
\end{split}
\end{equation}

From Eq.~\eqref{finalentstate}, we can see that for the evolution $U_+$, the part of the quantum state in the round brackets is symmetric under an exchange of particles. Importantly, the number of terms in this symmetric part of the state is $2d-2$. The symmetric fraction in the final state will then be $1-1/d$, and, consequently, in the limit as $d\longrightarrow \infty$, the state becomes almost fully symmetric. For the evolution $U_-$, the same effect can be seen, but here the component in the round brackets is anti-symmetric under an exchange of particles, such that the state becomes fully anti-symmetric in the limit $d\longrightarrow \infty$.

Hence, we can see how the transformations $U_{\pm}$ affect the symmetry of a state depending on its dimension $d$. Continuing on, the next step in the scheme is a symmetry projection. More specifically, the projector onto the anti-symmetric space $\Pi_A$ is given by:

\begin{equation} \label{antisymproj}
    \Pi_A=\mathbbm{1}-\Pi_S= \frac{1}{2} \left( \mathbbm{1}- S \right)\;,
\end{equation}

where $\Pi_S$ is the projector onto the symmetric space and $S= \sum_{i,j} \ket{i}_2\!\bra{j}\otimes\ket{j}_1\!\bra{i}$ is the swap operator.

\begin{figure}[h!]
    \includegraphics[width=8cm]{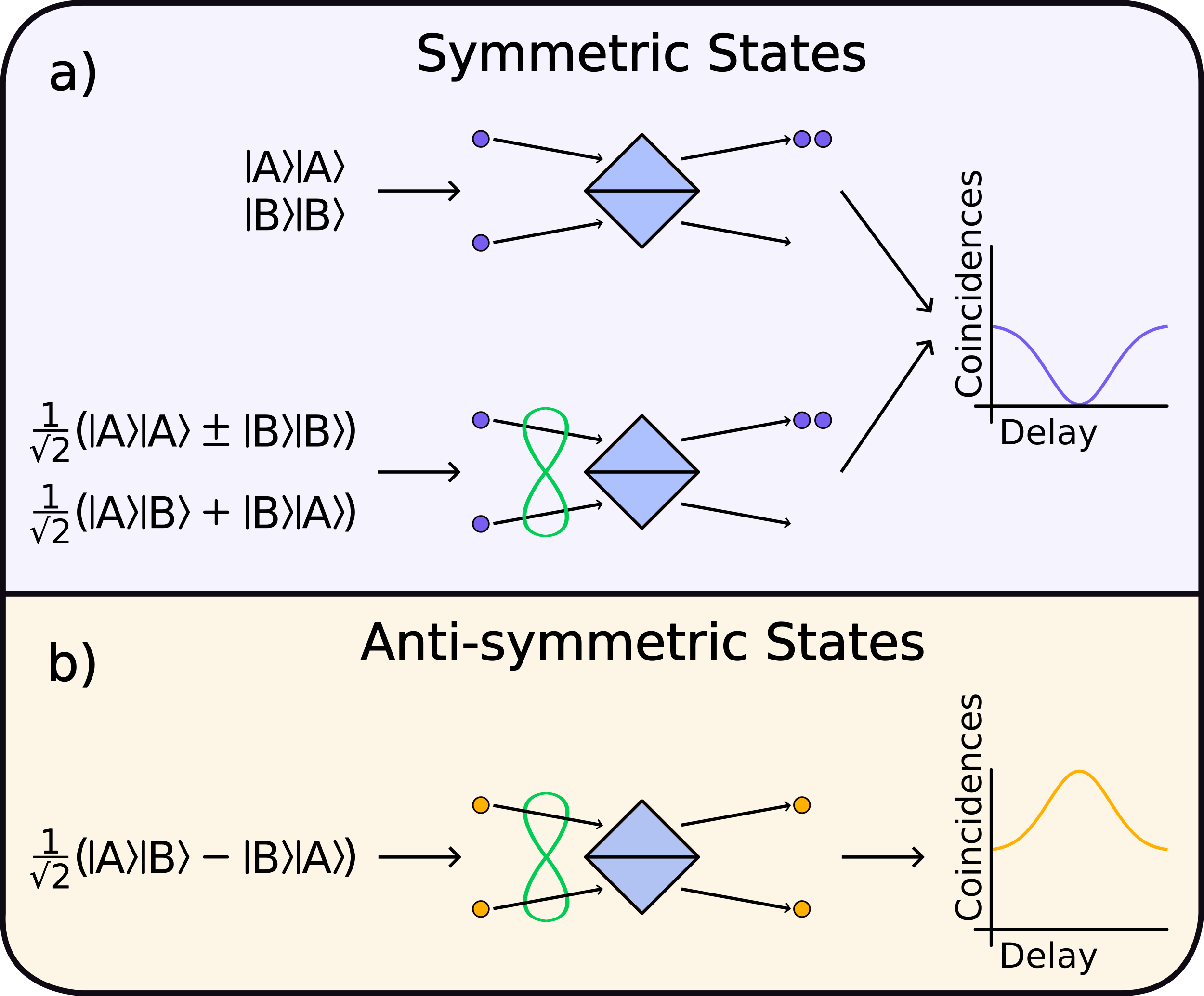}
    \centering
    \caption{\textbf{Dependence of the HOM effect on particle exchange symmetry.} \textbf{a)} If a two-particle state is symmetric under an exchange of particles, bunching occurs at the output of the HOM beamsplitter when these two particles are interfered. \textbf{b)} Conversely, if this state is anti-symmetric, anti-bunching occurs at the outputs. }
    \label{fig:symmetrytheory}
\end{figure}

The expectation value of this anti-symmetric projector can be directly obtained via a HOM measurement. HOM interference \cite{bouchard2020two} has already proven to be a robust tool for probing time-bin-encoded quantum states \cite{white2025robust}. Here, the measured coincidences serve to reveal the symmetry of our evolved time-bin superpositions. Since photons are bosons, any complete, multi-photon wave-function must be symmetric under the exchange of any two identical photons. Therefore, any anti-symmetric component in one degree of freedom must be compensated for by another. For example, the anti-symmetry of the time-bin components can be compensated by the anti-symmetry of the spatial degree of freedom. As illustrated in Fig.~\ref{fig:symmetrytheory}, HOM interference is sensitive to the symmetry of the interfering mode of the states: Anti-symmetric states produce a coincidence peak, whereas symmetric states yield a dip. Therefore, the coincidence probabilities extracted from a HOM measurement can be used to directly estimate the expectation values $\langle\Pi_A\rangle$ and $\langle\Pi_S\rangle$, which in turn can be used to estimate the value $\langle S \rangle$ (see Appendix \ref{apdx:measurementtheory}). 

Combining the evolutions in Eq.~\eqref{symtransforms} with the final HOM-based symmetry projection, the two dichotomic measurements $S_a$ of the scheme on the state $\rho$ are described by $S_a= U_a^\dagger S U_a$, where $a \in \{+,-\}$. In the following section, we show how to use these measurements to certify the entanglement of $\rho$ and non-trivially lower-bound its entanglement dimensionality.

\section{Robust certification of the Schmidt rank}
\subsection{Approach for state certification}
\label{TheoryMethod}
The intuitive idea behind our certification scheme is to quantify the symmetry properties of the transformed states and determine the minimum Schmidt number compatible with the observed measurement outcomes. In general, this is a difficult task, but by employing suitable Schmidt-number criteria, the problem can be relaxed into a semidefinite program (SDP) that searches for positive semidefinite matrices that reproduce the observed data while satisfying the chosen Schmidt-number constraints. A natural starting point is the well-known relation between Schmidt number and the fidelity $F$ with a maximally entangled state of local dimension $d$, namely $F d \leq k$ \cite{Terhal_2000}, where $k$ denotes the Schmidt number. One could therefore attempt to derive a lower bound on the fidelity from the behavior of the unknown state under the symmetry transformations introduced above. However, since there exist $d^2$ maximally entangled states, incorporating all corresponding fidelity constraints into the SDP would lead to a prohibitively large optimization problem and poor scaling with dimension. This observation motivated the construction of a new Schmidt-number witness that simultaneously captures a set of $d$ maximally entangled states:

\begin{equation}
    W(\rho) = \frac{1}{d} \sum_{i,j=0}^{d-1} |\bra{ii}\rho\ket{jj}| \; \leq \frac{k}{d}\;,
\end{equation} 

where the subscripts are omitted. Intuitively, this witness certifies proximity to the best-fitting state among a whole family of maximally entangled states, $\ket{\Phi_n^+}:= \frac{1}{\sqrt{d}} \sum_{k=0}^{d-1} e^{\frac{2\pi i}{d}kn} \ket{kk}$ for $n\in\{0,...,d-1\}$, including the state from Eq.~\eqref{targetstate}, thereby dramatically improving the scaling with dimension. A full proof of $W(\rho) \leq \frac{k}{d}$ is provided in Appendix \ref{apdx:derivationWitness}. We construct an optimization problem that minimizes our witness subject to linear constraints imposed by the observed behavior of the unknown state under the two symmetry-based measurements with outcome probabilities $C_{\textrm{exp, a}}$, $a \in \{\mathrm{+}, \mathrm{-}\}$, for measurement settings '+' and '-', represented by measurement operators $S_a$ (for a detailed definition, see Appendix \ref{apdx:SDP_Derivation}), respectively. This approach can be naturally relaxed to incorporate measurement uncertainty in its formulation; we denote lower and upper bounds on the outcome probabilities for both measurement setups by $C_{\textrm{exp, a}}^{\ell}$ and $C_{\textrm{exp, a}}^{u}$, respectively.

We note that, in the formalization of the evolution, the discretization is chosen before the symmetry transformations are applied, resulting in a Hilbert space of fixed dimension $d$. However, the applied delay can also populate a bin outside the original $d$-dimensional Hilbert space, thereby affecting the HOM measurement probabilities. In this work, we truncate the outer terms and renormalize the remaining state accordingly. Since this enlarges the set of states compatible with our observations, this will lead to a conservative bound on the entanglement dimension in the scenario we consider.

Furthermore, the fundamental properties of quantum states, such as Hermiticity, positive semidefiniteness, and the trace equaling one, can be directly phrased as constraints to the optimization problem. Beyond this however, one can optionally introduce assumptions by considering the properties that would arise due to the physical generation processes of the state. Here, we incorporate such properties in terms of linear operator equalities:

First, all entangled photon pairs produced via CW laser pumping experience the same temporal bias in their arrival times, hence are temporally uniform. We refer to this feature as the 'time-invariance property'. Mathematically, this is reflected in a block-wise symmetry, which implies that certain $(d+2) \times (d+2)$ diagonal principal submatrices of the state's density matrix must be identical. To enforce this, we introduce a family of linear maps $\mathcal{T}_n(X), n \in \{1,...,d-1\}$. Additional details and implementation aspects are provided in Appendix \ref{apdx:TimeInv}.

Second, the photon-pair-production process is fundamentally uncorrelated with noise photons. This independence gives rise to a symmetry among specific diagonal elements of the density matrix, which motivates the name 'diagonal symmetry property'. We express this symmetry using two families of linear maps $\mathcal{Q}^1_{i}(X), ~i\in\{1,...,d-1\}$ and $\mathcal{Q}^2_{i,j}(\rho), ~i,j \in \{1,...,d-1\}$. Further details are provided in Appendix \ref{apdx:SymmDiag}. 

Both properties arise directly from the stochastic nature of the photon-pair generation process under CW pumping. Consequently, incorporating them as optional constraints is well justified in non-adversarial scenarios, such as source characterization. Along with the constraints on the unknown quantum state imposed by the measurements, this leads to the following optimization problem:

\begin{equation}
\label{eq:optimizationProblem}
    \begin{aligned}
    \mathrm{minimise  }&~~ \frac{1}{d}\sum_{m,n=0}^{d-1} | \bra{mm}\rho\ket{nn}|\\
    \mathrm{s.t.: }&\\
    &C_{\mathrm{exp, a}}^{u} \geq \Tr{\rho\, S_{\mathrm{a}}} \geq  C_{\mathrm{exp, a}}^{\ell}  ~~~\forall a \in \{\mathrm{+}, \mathrm{-}\} \\ 
    &\Tr{\rho} -1 = 0 \\
    & \mathcal{T}_n(\rho) = 0 ~~~ \forall n \in \{1,..., d-1\} \\
    & \mathcal{Q}^1_{i}(\rho) = 0 ~~~ \forall i \in \{1,..., d-1\} \\
    & \mathcal{Q}^2_{i,j}(\rho) = 0 ~~~ \forall i \ne j \in \{0,...,d-1\} \\
    & \rho \geq 0. 
\end{aligned}
\end{equation}

\begin{figure*}
    \includegraphics[width=16cm]{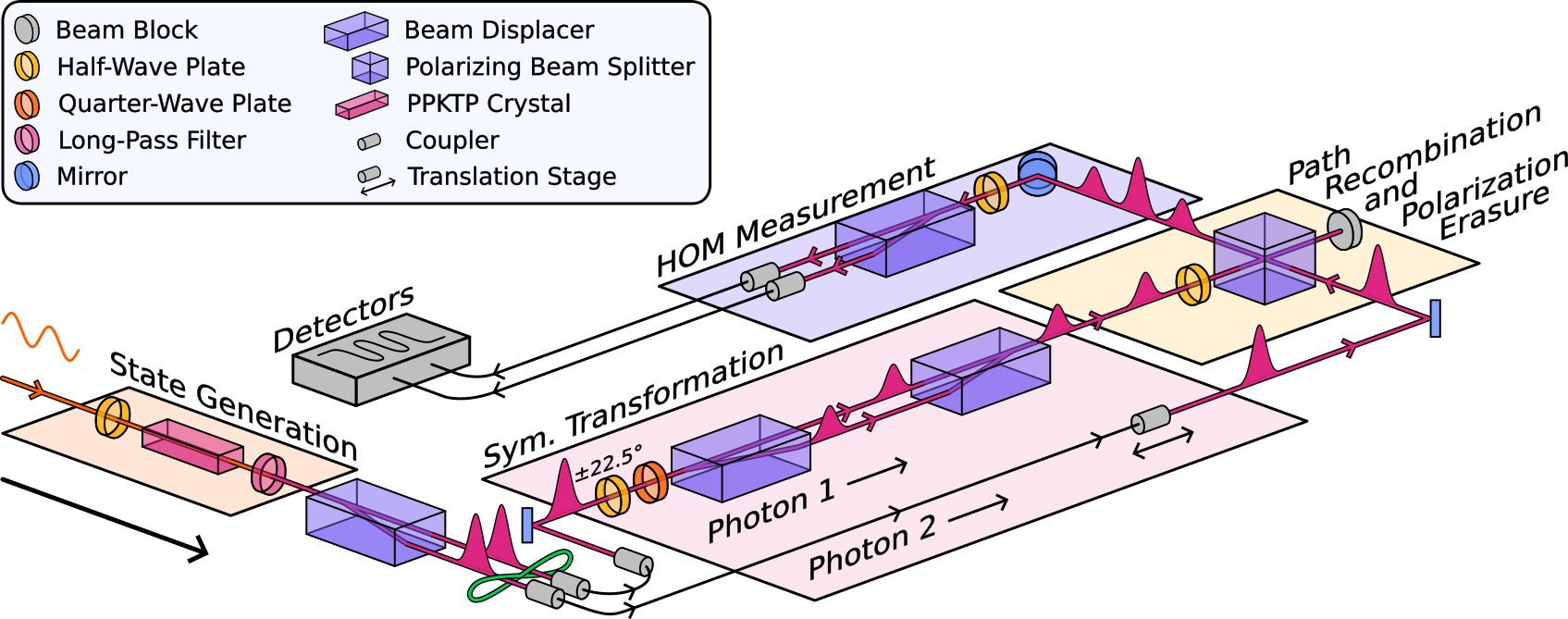}
    \centering
    \caption{\textbf{Experimental setup.} A type-II SPDC process is pumped by a 775 nm CW laser to generate a photon pair in the state described by Eq.~\eqref{mixedstate} (orange panel). This pair is deterministically separated based on polarization and coupled to two separate paths (red panel). One photon in this pair is passed through a polarization-based unbalanced interferometer, where the two different transformations $U_\pm$ in Eq.~\eqref{symtransforms} are selected by rotating a half-wave-plate at the beginning of the polarization Mach-Zehnder interferometer. The single-bin delay of $U_\pm$ on the other photon is implemented by a translation stage. Both photons are then passed through a polarizing beam splitter (yellow panel) to recombine their paths and erase the polarization information imparted by the interferometer on the first photon. This process completes the transformations $U_\pm$. Information on the time-bin state of the photons is then extracted from the results obtained via a polarization-based HOM measurement (purple panel). The coincidences for the HOM measurement are recorded using superconducting nanowire single-photon detectors (SNSPDs) and counting modules.}
    \label{fig:Apparatus}
\end{figure*}

This optimization problem can be formulated as an SDP that can be solved numerically. However, this comes with two disadvantages. First, the cost of solving this problem numerically scales extremely unfavorably with dimension, which restricts our analysis essentially to low double-digit dimensions. Second, found numerical solutions to this problem represent only approximations to the true minimum, which is not sufficient to reliably certify Schmidt numbers. Thus, we transform this primal SDP into its dual, which tackles both shortcomings simultaneously. The dual is computationally significantly less demanding, unlocking the analysis of higher-dimensional systems, and it provides us with reliable lower bounds on the Schmidt-number witness value.  We refer the reader to Appendix \ref{apdx:TransformationDual} for the derivation of the dual. The obtained dual optimization problem is then modeled in YALMIP \cite{Lofberg2004} and solved numerically using MOSEK 9.1.9 \cite{mosek}. In summary, our technique allows us to certify robust lower bounds on the dimension of the unknown quantum states provided, based on experimental observations. We note that the modulus-based witness was jointly proven in a concurrent project on time-bin entanglement \cite{schiffer2026brightsourcehighdimensionaltemporal}, whereas its integration into a dual of our SDP is unique to this work and is key to unlocking manageable scaling of computational resources. Having motivated and stated the relevant optimization problem, let us now discuss the physical properties of the analyzed quantum system in more detail.

\section{Experimental demonstration}
To experimentally demonstrate the scheme for the case of time-bin-encoding, we begin by generating pairs of degenerate 1550 nm photons with a state that in the ideal case is represented by Eq.~\eqref{targetstate} from an SPDC source pumped by a 775 nm CW laser. The two photons are then separated onto two paths, one containing a polarization-based unbalanced interferometer realized using two beam displacers, and the other a delay line (see Fig.~\ref{fig:Apparatus}). In this section of the experiment, either transformation of Eq.~\eqref{symtransforms} can be applied to the biphoton state.

Specifically, the transformation $U_-$ from Eq.~\eqref{symtransforms} is implemented by preparing the first photon in an anti-diagonal polarization state $\ket{-}=(\ket{H}-\ket{V})/\sqrt{2}$. Through the polarization-based unbalanced interferometer, the arrival time of the photon is coupled with its polarization such that each term $\ket{-, \; i}_1$ is transformed into a superposition $(\ket{H, \; i}_1-\ket{V, \; i+2}_1)/\sqrt{2}$. When the polarization information is later probabilistically erased by means of a half-wave plate (HWP) followed by a polarizing beam splitter (PBS), the transformation on the first photon $\ket{i}_1 \rightarrow (\ket{i}_1 - \ket{i+2}_1)/\sqrt{2}$ in the time-bin space is implemented. Likewise, the transformation $U_+$ can be realized in a similar way. In that case, the first photon is prepared in a diagonal polarization state $\ket{+}=(\ket{H}+\ket{V})/\sqrt{2}$, such that the interferometer in combination with the polarization erasure applies the following transformation in the time-bin domain: $\ket{i}_1 \rightarrow (\ket{i}_1 + \ket{i+2}_1)/\sqrt{2}$. In both cases, to complete these respective transformations, the second photon is passed through a delay line that imparts the necessary single-bin delay of $U_\pm$.

\begin{figure}
    \includegraphics[width=8.5cm]{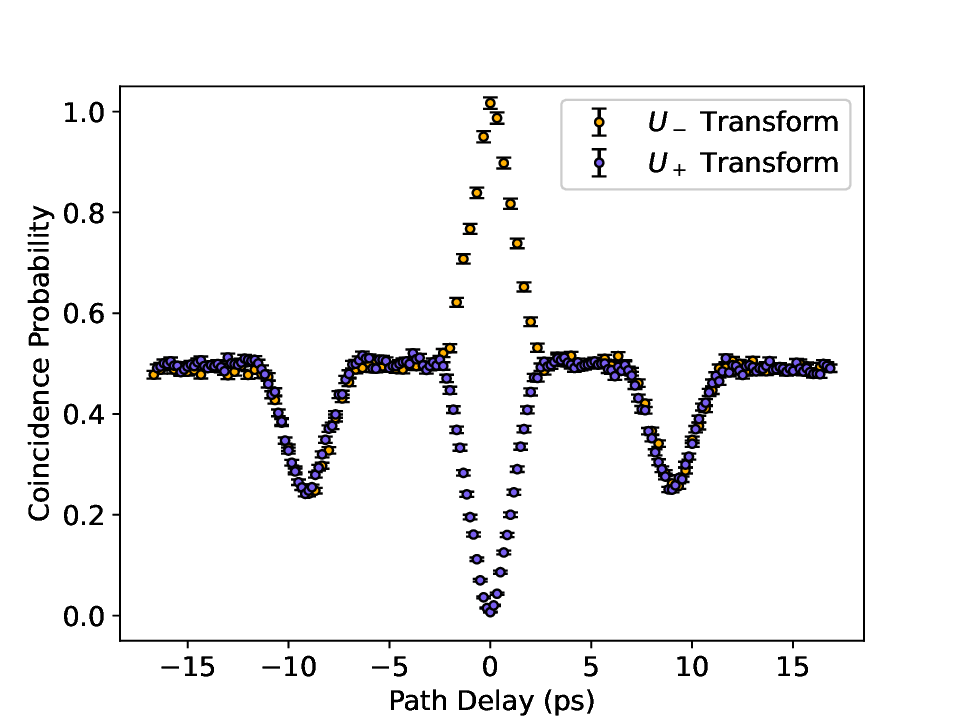}
    \centering
    \caption{\textbf{HOM scans.} Examples of the symmetric and anti-symmetric interference features, associated with the $U_\pm$ transforms, that arise as the two photons interfere during a HOM scan of relative path delay between the photons. The photons in this example plot were filtered by a 3 nm bandwidth filter centered at 1550 nm. The error bars here estimate the Poissonian fluctuation in the coincidence statistics and are too small to be clearly visible compared to the data points themselves.
    }
    \label{fig:HOMScans}
\end{figure}

After the transformation, the first photon is horizontally polarized, while the second is vertically polarized. To finally project our transformed state $\rho_{\pm}$ into the anti-symmetric space, we perform HOM interference in the polarization degree of freedom. This is achieved through the combination of a HWP and a beam displacer. The coincidence counts after the beam displacer are recorded for two HWP settings. First, the HWP is set to $0^\circ$ such that the photons always anti-bunch at the beam displacer, allowing us to estimate the total number of photon pairs $N^{\pm}$ entering the HOM measurement. Following this, the HWP is rotated by $22.5^\circ$ to implement the HOM measurement.  From these measured coincidence counts $\mathrm{CC}_{\text{HOM}}^{\pm}$ and $N^{\pm}$, we can retrieve the mean value of the projector $\Pi_A$ in Eq.~\eqref{antisymproj} applied to the evolved two-photon state $\rho_{\pm}$ incident on the HOM measurement setup: $\Tr\{\rho_{\pm} \Pi_A\}=C_{\mathrm{exp, \pm}}=\mathrm{CC}_{\text{HOM}}^{\pm}/N^{\pm}$. Likewise, we can estimate the visibility of the HOM features as $V=| \mathrm{CC}_{\text{HOM}}^{\pm}-N^{\pm}/2 |/(N^{\pm}/2)$, where the absolute value is taken in the numerator, and $N^{\pm}/2$ is equivalent to the standard baseline of the HOM dip. Traditionally, a HOM measurement on this state would be performed by scanning the delay of the second photon relative to the first, to build an interference picture of the whole state as in Fig. \ref{fig:HOMScans}. By instead fixing the delays of both photons and measuring only the two quantities $N^{\pm}$ and $\mathrm{CC}_{\text{HOM}}^{\pm}$, we avoid issues coming from possible misalignment during the delay scan. 

In our experiment, we obtain the following HOM visibilities at the final beam displacer, using a 1 nm filter to improve the spectral indistinguishability of the photons. We first apply the $U_-$ transformation in Eq.~\eqref{symtransforms}. Here, the almost (depending on the dimension $d$) anti-symmetric two-photon state anti-bunches, and we obtain a visibility of $V^{\mathrm{exp}}_{\mathrm{peak}}= 0.990 \pm 0.004$. Next, we apply the $U_+$ transformation in Eq.~\eqref{symtransforms}, such that the photons bunch at the output of the beam displacer, and we obtain a visibility of $V^{\mathrm{exp}}_{\mathrm{dip}}= 0.9898 \pm 0.0002$. In both of these cases, the errors in visibility are determined by assuming Poissonian fluctuations in the observed coincidence counts and propagating them accordingly. In the next section, we use these measurements to certify the dimensionality of an unknown initial quantum state.

\section{Results} \label{sec:results}
To certify entanglement and its dimensionality from the observed data, one may impose different levels of assumptions on the source state. Stronger assumptions, such as the purity of the unknown quantum state, reflect a higher degree of trust in the source and lead to higher certified entanglement dimensions. Here, however, we aim to eliminate assumptions on the knowledge of the source state, trusting only the measurement devices and their independence from the source. Accordingly, we analyze two scenarios: a non-adversarial setting, in which properties of the source and its relation with noises are taken into account, and an adversarial scenario in which no properties of the source are considered.

\begin{figure}
    \includegraphics[width=7cm]{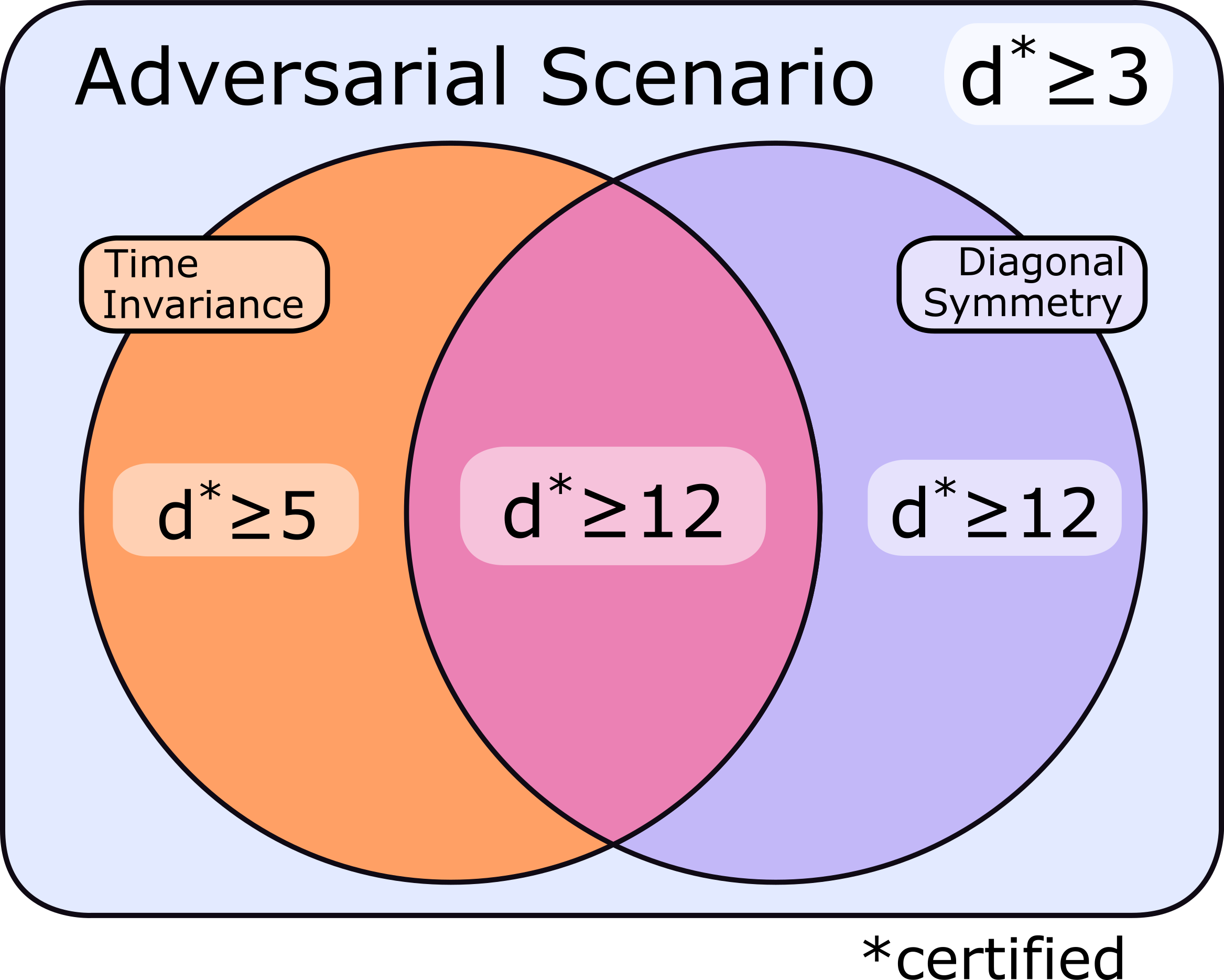}
    \centering
    \caption{\textbf{Entanglement dimension certification.} The minimum number of dimensions certified from the HOM data for different assumptions considered. Each lower bound is calculated from the worst-case scenario compatible with the experimental error.}
    \label{fig:VennDiagram}
\end{figure}

\subsection{Robust certification of entanglement dimension bounds}
Although considering the state to be pure would allow us to certify a very high degree of entanglement within the state without needing the SDP technique (see Appendix \ref{apdx:maxent}), this approach is extremely restricted in its generality and is inappropriate for the analysis of the unknown, arbitrary states we consider. Instead, using the measured experimental probabilities and our developed optimization program in Eq.~\eqref{eq:optimizationProblem}, we can retrieve lower bounds on the dimensionality of the state entanglement without restrictive assumptions on the state. We consider the two different properties rooted in the state generation and outlined prior to Eq.~\eqref{eq:optimizationProblem}, that is, we assume the photon generation process is uniform in time (time-invariance), and we assume the noises affecting the state are uncorrelated with the generation process (diagonal symmetry). If we consider only the time-invariance, we are able to certify a minimum entanglement dimension of $d\ge5$. Likewise, if we only consider the diagonal symmetry, we are able to certify a minimum entanglement dimension of $d\ge12$. If we consider both of these properties simultaneously, we still certify a minimum entanglement dimension of $d\ge12$; however, this bound on the entanglement dimension is certified with a lower modeled state dimension, as seen in Appendix \ref{apdx:CertifiedEntanglementDimension}.

\subsection{Adversarial scenario}
Finally, we also consider the case without any assumptions on the initial state in Eq.~\eqref{mixedstate} and, here, notably, we still obtain a nontrivial minimum dimension of $d\ge3$. This demonstrates the power of our approach: Two symmetry-based dichotomic measurements suffice to certify entanglement in the high-dimensional regime even in the adversarial scenario.

\section{Discussion}
Previously, HOM interference was only known as a tool for observing the \emph{presence} of entanglement. Here, we have shown how HOM interference can be combined with linear-optical symmetry manipulations to go much further and not only detect but even \emph{quantify} entanglement in the high-dimensional realm.
We develop a simple scheme that certifies a lower bound on the entanglement dimensionality of time-bin-encoded quantum states using only two symmetry-based, dichotomic measurements, with minimal assumptions about the underlying quantum state. We experimentally demonstrate this scheme using the example of biphoton states produced by a CW-pumped SPDC source. From this, we certify an entanglement dimensionality of 12 with only the natural physical properties of time-invariance and diagonal symmetry. Strikingly, even when we do not take those properties into account, the scheme continues to certify a nontrivial 3-dimensional entanglement. We highlight that our observations are compatible with entanglement dimensions that are significantly larger; however, in certification, the task is to certify a certain entanglement dimension in the worst-case non-adversarial scenario, without assumptions on the state itself. With a less conservative approach---for example, assuming purity---we could certify entanglement with triple-digit dimensions (see Appendix \ref{apdx:maxent}). 

Alternative approaches have also been developed to certify high-dimensional entanglement across a range of photonic experiments; however, they typically involve measurement schemes whose number of measurements and/or outcomes grows with dimension~\cite{Nape2021, HerreraValencia2020highdimensional, Ndagano2020, PRXQuantum.4.010308, MarioKrenn2014, Bavaresco2018, schiffer2026brightsourcehighdimensionaltemporal, ponce2023, xiaomin2020} and some rely on restrictive assumptions, such as the underlying state being pure~\cite{Chang2021, Cheng2023, Courme2023}. Compared to these approaches, our approach offers important strengths. 

Firstly, our measurement scheme is independent of the state dimensionality and requires only two dichotomic measurements. This is in stark contrast to the linear and quadratic scalings of the approaches above and to tomographic methods, whose complexity grows exponentially with the dimension. 

Secondly, our approach operates at any timescale, even at temporal resolutions far below those of the detectors and the coincidence logic. Indeed, in our experiment, both the coherence length of our photons ($\sim4.3$~ps) and time-bin separation ($\sim9.1$~ps) are already below the temporal resolution ($>400$~ps) of our detectors (due to jitter), and are also near the resolution limit of state-of-the-art superconducting detector jitter and coincidence windows~\cite{korzh2020demonstration,esmaeil2020efficient}. Fundamentally, there is no reason the choice of time-bin discretization needs to match the delays in the setup as specified in Eq.~(\ref{finalentstate}). Indeed, the same strategy of symmetry manipulations allows us to move to even smaller timescales by simply redefining the time-bin size during the data analysis. Although this quickly becomes computationally challenging, we still managed to observe high-dimensional entanglement while reducing the time-bin size by a factor of 4, using the same data. This freedom to change discretization opens the possibility of going far below any timescale resolvable by state-of-the-art time-tagging technology and revealing entanglement even at ultra-short timescales (see Appendix \ref{apdx:CertifiedEntanglementDimension} for details).

Finally, in addition to avoiding demanding temporal detection-resolution requirements, our approach also eliminates the need for complex experimental setups and strong theoretical assumptions commonly encountered in alternative methods for characterizing the entanglement of high-dimensional time-bin states. 

While offering the above benefits, our scheme also operates within certain constraints. First, it applies to bipartite states, with no immediate, obvious way to extend the approach to multipartite states. Second, as with other HOM-based schemes, our scheme involves a local measurement, since the two photons interfere. Hence, it is best to use the scheme as part of a source characterization before moving on to an application in which the photons from the source are sent to spatially separated parties. Third, the dimensions certified by our method can be limited by experimental imperfections, such as photon distinguishability (which degrades HOM visibility) and uncertainty in the number of coincidences, in combination with the SDP analysis. We note that we obtain our certified dimensionalities \emph{without} correcting for imperfections in the indistinguishability or similar factors limiting the baseline HOM visibility. For comparison, if we had corrected for the baseline HOM visibility (measured without the transformations applied), the certified entanglement dimension, using diagonal symmetry and time-invariance, would have been 13 rather than 12. From this, we conclude that photon distinguishability alone is not a significant limitation on our certified dimension.

Finally, for the analysis, we model the interferometric transformations and the corresponding measurement operators in a discretized time-bin description, introducing a $d$-dimensional computational space per input photon and the joint space $\mathcal{H}_d=\mathcal H_d^{(A)}\otimes \mathcal H_d^{(B)}$ with $\mathcal H_d^{(X)}=\mathrm{span}\{\ket{0},\ldots,\ket{d{-}1}\}$. A standard subtlety in time-bin implementations is that superposition measurements realized via an interferometric delay need not preserve this initial space: Near the boundary of the modeled space, the photon's delayed component can populate bins outside $\mathcal H_d$ and can affect HOM-based measurement probabilities. In the SDP-based relaxation, rather than modeling these additional modes explicitly (see, e.g.~\cite{kanitschar2024practical}), we deliberately truncate the evolved state back onto $\mathcal H_d$ and suitably renormalize. This step allows us to impose symmetries and substantially simplifies the semidefinite relaxation, while still bounding the statistics accessible by $d$-dimensional states. The renormalization requires that the delay acts equivalently on all elements of the computational basis, which is physically well-justified. However, by discarding out-of-subspace structure, we reduce the predictive power of the description and thereby enlarge the set of states compatible with the observed HOM statistics. Consequently, any entanglement dimension certified by the SDP holds and is conservative, yet still allows us to certify nontrivial high-dimensional entanglement. 

Beyond its fundamental significance, our conceptually simple and experimentally accessible approach provides a practical framework for characterizing high-dimensional time-bin entangled states. By enabling certification with only two dichotomic measurements, it greatly reduces the overhead usually associated with high-dimensional entanglement certification.
This makes it particularly well-suited to scenarios in which full state tomography is prohibitively costly, but reliable and scalable verification remains essential. Prominent examples include source qualification in quantum networks, certification of high-dimensional sources for quantum communication \cite{xavier2025energy}, and integrated photonic time-bin platforms \cite{montaut2025progress}, where compactness and low experimental complexity are especially valuable. The same features also make the method attractive for entanglement verification in long-distance and repeater-based links \cite{azuma2023quantum}, where efficient monitoring of high-dimensional states before their distribution is a central challenge. More broadly, the scheme is not restricted to time-bin encoding: In principle, it can be extended to other high-dimensional photonic degrees of freedom and to alternative qudit platforms, such as spin systems and atomic ensembles, provided they support the required symmetry operations and HOM measurements.

\begin{acknowledgments}
N.T.~was a recipient of an Australian Research Council Discovery Early Career Researcher Award (DE220101082), and  E.P.~is a recipient of an Australian Research Council Discovery Early Career Researcher Award (DE250100762). F.K.~and M.H.~acknowledge funding from the Horizon-Europe research and innovation programme under grant agreement No 101070168 (HyperSpace) and from the European Union{\textemdash}NextGenerationEU project FO999921415 (Vanessa-QC). F.K.~gratefully acknowledges support from the Dieberger-Skalicky foundation.
\begin{figure}[htb!]
\centering
\includegraphics[width=0.4\columnwidth]{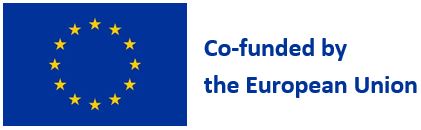}\end{figure}
\end{acknowledgments}

\bibliography{bibliography}

@article{horodecki2009quantum,
  title={Quantum entanglement},
  author={Horodecki, Ryszard and Horodecki, Pawe{\l} and Horodecki, Micha{\l} and Horodecki, Karol},
  journal={Reviews of modern physics},
  volume={81},
  number={2},
  pages={865--942},
  year={2009},
  publisher={APS},
doi={https://doi.org/10.1103/RevModPhys.81.865}
}

@article{brunner2014bell,
  title={Bell nonlocality},
  author={Brunner, Nicolas and Cavalcanti, Daniel and Pironio, Stefano and Scarani, Valerio and Wehner, Stephanie},
  journal={Reviews of modern physics},
  volume={86},
  number={2},
  pages={419--478},
  year={2014},
  publisher={APS},
doi={https://doi.org/10.1103/RevModPhys.86.419}
}

@article{azuma2023quantum,
  title={Quantum repeaters: From quantum networks to the quantum internet},
  author={Azuma, Koji and Economou, Sophia E and Elkouss, David and Hilaire, Paul and Jiang, Liang and Lo, Hoi-Kwong and Tzitrin, Ilan},
  journal={Reviews of Modern Physics},
  volume={95},
  number={4},
  pages={045006},
  year={2023},
  publisher={APS},
  doi={https://doi.org/10.1103/RevModPhys.95.045006}
}

@article{xu2020secure,
  title={Secure quantum key distribution with realistic devices},
  author={Xu, Feihu and Ma, Xiongfeng and Zhang, Qiang and Lo, Hoi-Kwong and Pan, Jian-Wei},
  journal={Reviews of modern physics},
  volume={92},
  number={2},
  pages={025002},
  year={2020},
  publisher={APS},
doi={https://doi.org/10.1103/RevModPhys.92.025002}
}

@article{friis2019entanglement,
  title={Entanglement certification from theory to experiment},
  author={Friis, Nicolai and Vitagliano, Giuseppe and Malik, Mehul and Huber, Marcus},
  journal={Nature Reviews Physics},
  volume={1},
  number={1},
  pages={72--87},
  year={2019},
  publisher={Nature Publishing Group UK London},
doi={https://doi.org/10.1038/s42254-018-0003-5}
}

@article{wang2020qudits,
  title={Qudits and high-dimensional quantum computing},
  author={Wang, Yuchen and Hu, Zixuan and Sanders, Barry C and Kais, Sabre},
  journal={Frontiers in Physics},
  volume={8},
  pages={589504},
  year={2020},
  publisher={Frontiers Media SA},
doi={https://doi.org/10.3389/fphy.2020.589504}
}

@article{graffitti2020direct,
  title={Direct generation of tailored pulse-mode entanglement},
  author={Graffitti, Francesco and Barrow, Peter and Pickston, Alexander and Bra{\'n}czyk, Agata M and Fedrizzi, Alessandro},
  journal={Physical Review Letters},
  volume={124},
  number={5},
  pages={053603},
  year={2020},
  publisher={APS},
  url={https://doi.org/10.1103/PhysRevLett.124.053603}
}

@article{barbieri2017hong,
  title={What Hong-Ou-Mandel interference says on two-photon frequency entanglement},
  author={Barbieri, Marco and Roccia, Emanuele and Mancino, Luca and Sbroscia, Marco and Gianani, Ilaria and Sciarrino, Fabio},
  journal={Scientific reports},
  volume={7},
  number={1},
  pages={7247},
  year={2017},
  publisher={Nature Publishing Group UK London},
  url={https://doi.org/10.1038/s41598-017-07555-4}
}

@article{eckstein2008broadband,
  title={Broadband frequency mode entanglement in waveguided parametric downconversion},
  author={Eckstein, Andreas and Silberhorn, Christine},
  journal={Optics letters},
  volume={33},
  number={16},
  pages={1825--1827},
  year={2008},
  publisher={Optical Society of America},
  url={https://doi.org/10.1364/OL.33.001825}
}

@article{kaneda2019direct,
  title={Direct generation of frequency-bin entangled photons via two-period quasi-phase-matched parametric downconversion},
  author={Kaneda, Fumihiro and Suzuki, Hirofumi and Shimizu, Ryosuke and Edamatsu, Keiichi},
  journal={Optics express},
  volume={27},
  number={2},
  pages={1416--1424},
  year={2019},
  publisher={Optical Society of America},
  url={https://doi.org/10.1364/OE.27.001416}
}

@article{maltese2020generation,
  title={Generation and symmetry control of quantum frequency combs},
  author={Maltese, G and Amanti, MI and Appas, F{\'e}licien and Sinnl, G and Lema{\^\i}tre, A and Milman, P and Baboux, F and Ducci, S},
  journal={npj Quantum Information},
  volume={6},
  number={1},
  pages={13},
  year={2020},
  publisher={Nature Publishing Group UK London},
  url={https://doi.org/10.1038/s41534-019-0237-9}
}

@article{merkouche2022heralding,
  title={Heralding multiple photonic pulsed bell pairs via frequency-resolved entanglement swapping},
  author={Merkouche, Sofiane and Thiel, Val{\'e}rian and Davis, Alex OC and Smith, Brian J},
  journal={Physical Review Letters},
  volume={128},
  number={6},
  pages={063602},
  year={2022},
  publisher={APS},
  url={https://doi.org/10.1103/PhysRevLett.128.063602}
}

@article{gao2022manipulating,
  title={Manipulating the symmetry of transverse momentum entangled biphoton states},
  author={Gao, Xiaoqin and Zhang, Yingwen and D’Errico, Alessio and Hufnagel, Felix and Heshami, Khabat and Karimi, Ebrahim},
  journal={Optics Express},
  volume={30},
  number={12},
  pages={21276--21281},
  year={2022},
  publisher={Optica Publishing Group},
  url={https://doi.org/10.1364/OE.458776}
}

@article{chen2022telecom,
  title={Telecom-band hyperentangled photon pairs from a fiber-based source},
  author={Chen, Changjia and Xu, Calvin and Riazi, Arash and Zhu, Eric Y and Greenwood, Alexander CB and Gladyshev, Alexey V and Kazansky, Peter G and Kirby, Brian T and Qian, Li},
  journal={Physical Review A},
  volume={105},
  number={4},
  pages={043702},
  year={2022},
  publisher={APS},
  url={https://doi.org/10.1103/PhysRevA.105.043702}
}

@article{chen2020verification,
  title={Verification of high-dimensional entanglement generated in quantum interference},
  author={Chen, Yuanyuan and Ecker, Sebastian and Bavaresco, Jessica and Scheidl, Thomas and Chen, Lixiang and Steinlechner, Fabian and Huber, Marcus and Ursin, Rupert},
  journal={Physical Review A},
  volume={101},
  number={3},
  pages={032302},
  year={2020},
  publisher={APS},
  url={https://doi.org/10.1103/PhysRevA.101.032302}
}

@article{chen2021temporal,
  title={Temporal distinguishability in Hong-Ou-Mandel interference for harnessing high-dimensional frequency entanglement},
  author={Chen, Yuanyuan and Ecker, Sebastian and Chen, Lixiang and Steinlechner, Fabian and Huber, Marcus and Ursin, Rupert},
  journal={npj Quantum Information},
  volume={7},
  number={1},
  pages={167},
  year={2021},
  publisher={Nature Publishing Group UK London},
  url={https://doi.org/10.1038/s41534-021-00504-0}
}

@article{polino2020photonic,
  title={Photonic quantum metrology},
  author={Polino, Emanuele and Valeri, Mauro and Spagnolo, Nicol{\`o} and Sciarrino, Fabio},
  journal={AVS Quantum Science},
  volume={2},
  number={2},
  year={2020},
  publisher={AIP Publishing},
url={https://doi.org/10.1116/5.0007577}
}

@article{white2025robust,
  title={Robust Approach for Time-Bin-Encoded Photonic Quantum Information Protocols},
  author={White, Simon JU and Polino, Emanuele and Ghafari, Farzad and Joch, Dominick J and Villegas-Aguilar, Luis and Shalm, Lynden K and Verma, Varun B and Huber, Marcus and Tischler, Nora},
  journal={Physical Review Letters},
    doi={https://doi.org/10.1103/PhysRevLett.134.180802},
  volume={134},
  number={18},
  pages={180802},
  year={2025},
  publisher={APS}
}

@book{rovelli2019order,
  title={The order of time},
  author={Rovelli, Carlo},
  year={2019},
  publisher={Penguin},
  url={https://www.penguin.com.au/books/the-order-of-time-9780141984964}
}

@article{bouchard2022quantum,
  title={Quantum communication with ultrafast time-bin qubits},
  author={Bouchard, Fr{\'e}d{\'e}ric and England, Duncan and Bustard, Philip J and Heshami, Khabat and Sussman, Benjamin},
  journal={PRX Quantum},
  volume={3},
  number={1},
  pages={010332},
  year={2022},
  publisher={APS},
doi={https://doi.org/10.1103/PRXQuantum.3.010332}
}

@article{ikuta2022scalable,
  title={Scalable implementation of (d+ 1) mutually unbiased bases for d-dimensional quantum key distribution},
  author={Ikuta, Takuya and Akibue, Seiseki and Yonezu, Yuya and Honjo, Toshimori and Takesue, Hiroki and Inoue, Kyo},
  journal={Physical Review Research},
  volume={4},
  number={4},
  pages={L042007},
  year={2022},
  publisher={APS},
  doi={https://doi.org/10.1103/PhysRevResearch.4.L042007}
}

@incollection{preskill2023quantum,
  title={Quantum computing 40 years later},
  author={Preskill, John},
  booktitle={Feynman Lectures on Computation},
  pages={193--244},
  year={2023},
  publisher={CRC Press},
  url = {https://www.amazon.science/publications/quantum-computing-40-years-later}
}

@book{muga2007time,
  title={Time in quantum mechanics},
  author={Muga, Gonzalo and Mayato, R Sala and Egusquiza, Inigo},
  volume={734},
  year={2007},
  publisher={Springer},
  url={https://link.springer.com/book/10.1007/978-3-540-73473-4}
}

@article{barbieri2022optical,
  title={Optical quantum metrology},
  author={Barbieri, Marco},
  journal={PRX Quantum},
  volume={3},
  number={1},
  pages={010202},
  year={2022},
  publisher={APS},
doi={https://doi.org/10.1103/PRXQuantum.3.010202}
}

@article{marcikic2003long,
  title={Long-distance teleportation of qubits at telecommunication wavelengths},
  author={Marcikic, Ivan and De Riedmatten, Hugues and Tittel, Wolfgang and Zbinden, Hugo and Gisin, Nicolas},
  journal={Nature},
  volume={421},
  number={6922},
  pages={509--513},
  year={2003},
doi={https://doi.org/10.1038/nature01376},
  publisher={Nature Publishing Group UK London}
}

@incollection{wiseman2017causarum,
  title={{Causarum Investigatio and the two Bell’s theorems of John Bell}},
  author={Wiseman, Howard M and Cavalcanti, Eric G},
  booktitle={Quantum [Un] Speakables II},
  pages={119--142},
  year={2017},
  publisher={Springer},
  url={https://doi.org/10.1007/978-3-319-38987-5_6}
}

@book{nielsen2010quantum,
  title={Quantum computation and quantum information},
  author={Nielsen, Michael A and Chuang, Isaac L},
  year={2010},
  publisher={Cambridge University Press},
  url={https://doi.org/10.1017/CBO9780511976667}
}

@article{kanitschar2024practical,
  title = {Practical Framework for Analyzing High-Dimensional Quantum Key Distribution Setups},
  author = {Kanitschar, Florian and Huber, Marcus},
  journal = {Phys. Rev. Lett.},
  volume = {135},
  issue = {1},
  pages = {010802},
  numpages = {6},
  year = {2025},
  month = {Jul},
  publisher = {American Physical Society},
  url = {https://link.aps.org/doi/10.1103/PhysRevLett.135.010802}
}

@article{takesue2009implementation,
  title={Implementation of quantum state tomography for time-bin entangled photon pairs},
  author={Takesue, Hiroki and Noguchi, Yuita},
  journal={Optics Express},
  volume={17},
  number={13},
  pages={10976--10989},
  year={2009},
doi={https://doi.org/10.1364/OE.17.010976},
  publisher={Optica Publishing Group}
}

@article{hong1987measurement,
  title={Measurement of subpicosecond time intervals between two photons by interference},
  author={Hong, Chong-Ki and Ou, Zhe-Yu and Mandel, Leonard},
  journal={Physical Review Letters},
  volume={59},
  number={18},
  pages={2044},
  year={1987},
doi={https://doi.org/10.1103/PhysRevLett.59.2044},
  publisher={APS}
}

@article{islam2017provably,
  title={Provably secure and high-rate quantum key distribution with time-bin qudits},
  author={Islam, Nurul T and Lim, Charles Ci Wen and Cahall, Clinton and Kim, Jungsang and Gauthier, Daniel J},
  journal={Science Advances},
  volume={3},
  number={11},
  pages={e1701491},
  year={2017},
doi={10.1126/sciadv.1701491},
  publisher={American Association for the Advancement of Science}
}

@article{wen2022realizing,
  title={Realizing an entanglement-based multiuser quantum network with integrated photonics},
  author={Wen, Wenjun and Chen, Zhiyu and Lu, Liangliang and Yan, Wenhan and Xue, Wenyi and Zhang, Peiyu and Lu, Yanqing and Zhu, Shining and Ma, Xiao-song},
  journal={Physical Review Applied},
  volume={18},
  number={2},
  pages={024059},
  year={2022},
doi={https://doi.org/10.1103/PhysRevApplied.18.024059},
  publisher={APS}
}

@article{kuzucu2008time,
  title={Time-resolved single-photon detection by femtosecond upconversion},
  author={Kuzucu, Onur and Wong, Franco NC and Kurimura, Sunao and Tovstonog, Sergey},
  journal={Optics Letters},
  volume={33},
  number={19},
  pages={2257--2259},
  year={2008},
doi={https://doi.org/10.1364/OL.33.002257},
  publisher={Optica Publishing Group}
}

@article{maclean2018direct,
  title={Direct characterization of ultrafast energy-time entangled photon pairs},
  author={MacLean, Jean-Philippe W and Donohue, John M and Resch, Kevin J},
  journal={Physical Review Letters},
  volume={120},
  number={5},
  pages={053601},
  year={2018},
doi={https://doi.org/10.1103/PhysRevLett.120.053601},
  publisher={APS}
}

@article{valivarthi2020teleportation,
  title={Teleportation systems toward a quantum internet},
  author={Valivarthi, Raju and Davis, Samantha I and Pe{\~n}a, Cristi{\'a}n and Xie, Si and Lauk, Nikolai and Narv{\'a}ez, Lautaro and Allmaras, Jason P and Beyer, Andrew D and Gim, Yewon and Hussein, Meraj and others},
  journal={PRX Quantum},
  volume={1},
  number={2},
  pages={020317},
  year={2020},
doi={https://doi.org/10.1103/PRXQuantum.1.020317},
  publisher={APS}
}

@article{zhang2014unconditional,
  title={Unconditional security of time-energy entanglement quantum key distribution using dual-basis interferometry},
  author={Zhang, Zheshen and Mower, Jacob and Englund, Dirk and Wong, Franco NC and Shapiro, Jeffrey H},
  journal={Physical review letters},
  volume={112},
  number={12},
  pages={120506},
  year={2014},
  publisher={APS},
  doi={https://doi.org/10.1103/PhysRevLett.112.120506}
}

@article{richart2012experimental,
  title={Experimental implementation of higher dimensional time--energy entanglement},
  author={Richart, Daniel and Fischer, Yvo and Weinfurter, Harald},
  journal={Applied Physics B},
  volume={106},
  number={3},
  pages={543--550},
  year={2012},
  publisher={Springer},
  doi={https://doi.org/10.1007/s00340-011-4854-z}
}

@article{zahidy2024practical,
  title={Practical high-dimensional quantum key distribution protocol over deployed multicore fiber},
  author={Zahidy, Mujtaba and Ribezzo, Domenico and De Lazzari, Claudia and Vagniluca, Ilaria and Biagi, Nicola and M{\"u}ller, Ronny and Occhipinti, Tommaso and Oxenl{\o}we, Leif K and Galili, Michael and Hayashi, Tetsuya and others},
  journal={Nature Communications},
  volume={15},
  number={1},
  pages={1651},
  year={2024},
  publisher={Nature Publishing Group UK London},
  doi={https://doi.org/10.1038/s41467-024-45876-x}
}

@article{zhu2021high,
  title={Is high-dimensional photonic entanglement robust to noise?},
  author={Zhu, Feng and Tyler, Max and Valencia, Natalia Herrera and Malik, Mehul and Leach, Jonathan},
  journal={AVS Quantum Science},
  volume={3},
  number={1},
  year={2021},
  publisher={AIP Publishing},
  url={https://doi.org/10.1116/5.0033889}
}

@article{yu2025quantum,
  title={Quantum key distribution implemented with d-level time-bin entangled photons},
  author={Yu, Hao and Sciara, Stefania and Chemnitz, Mario and Montaut, Nicola and Crockett, Benjamin and Fischer, Bennet and Helsten, Robin and Wetzel, Benjamin and Goebel, Thorsten A and Kr{\"a}mer, Ria G and others},
  journal={Nature Communications},
  volume={16},
  number={1},
  pages={171},
  year={2025},
  publisher={Nature Publishing Group UK London},
  doi={https://doi.org/10.1038/s41467-024-55345-0}
}

@article{bouchard2023measuring,
  title={Measuring ultrafast time-bin qudits},
  author={Bouchard, Fr{\'e}d{\'e}ric and Bonsma-Fisher, Kent and Heshami, Khabat and Bustard, Philip J and England, Duncan and Sussman, Benjamin},
  journal={Physical Review A},
  volume={107},
  number={2},
  pages={022618},
  year={2023},
doi={https://doi.org/10.1103/PhysRevA.107.022618},
  publisher={APS}
}

@article{erhard2020advances,
  title={Advances in high-dimensional quantum entanglement},
  author={Erhard, Manuel and Krenn, Mario and Zeilinger, Anton},
  journal={Nature Reviews Physics},
  volume={2},
  number={7},
  pages={365--381},
  year={2020},
doi={https://doi.org/10.1038/s42254-020-0193-5},
  publisher={Nature Publishing Group UK London}
}

@article{cozzolino2019high,
  title={High-dimensional quantum communication: benefits, progress, and future challenges},
  author={Cozzolino, Daniele and Da Lio, Beatrice and Bacco, Davide and Oxenl{\o}we, Leif Katsuo},
  journal={Advanced Quantum Technologies},
  volume={2},
  number={12},
  pages={1900038},
  year={2019},
doi={ https://doi.org/10.1002/qute.201900038},
  publisher={Wiley Online Library}
}

@article{fitzke2022scalable,
  title={Scalable network for simultaneous pairwise quantum key distribution via entanglement-based time-bin coding},
  author={Fitzke, Erik and Bialowons, Lucas and Dolejsky, Till and Tippmann, Maximilian and Nikiforov, Oleg and Walther, Thomas and Wissel, Felix and Gunkel, Matthias},
  journal={PRX Quantum},
  volume={3},
  number={2},
  pages={020341},
  year={2022},
doi={https://doi.org/10.1103/PRXQuantum.3.020341},
  publisher={APS}
}

@article{de2002creating,
  title={Creating high dimensional entanglement using mode-locked lasers},
  author={De Riedmatten, Hugues and Marcikic, Ivan and Zbinden, Hugo and Gisin, Nicolas},
  journal={Quantum Information \& Computation},
  volume={2},
  number={6},
  pages={425--433},
  year={2002},
doi={https://dl.acm.org/doi/10.5555/2011577.2011578},
  publisher={Rinton Press, Incorporated Paramus, NJ}
}

@article{ramelow2009discrete,
  title={Discrete tunable color entanglement},
  author={Ramelow, S and Ratschbacher, L and Fedrizzi, A and Langford, NK and Zeilinger, A},
  journal={Physical Review Letters},
  volume={103},
  number={25},
  pages={253601},
  year={2009},
  publisher={APS},
  url={https://doi.org/10.1103/PhysRevLett.103.253601}
}

@article{marcikic2004distribution,
  title={Distribution of time-bin entangled qubits over 50 km of optical fiber},
  author={Marcikic, Ivan and De Riedmatten, Hugues and Tittel, Wolfgang and Zbinden, Hugo and Legr{\'e}, Matthieu and Gisin, Nicolas},
  journal={Physical Review Letters},
  volume={93},
  number={18},
  pages={180502},
  year={2004},
doi={https://doi.org/10.1103/PhysRevLett.93.180502},
  publisher={APS}
}

@article{zhong2015photon,
  title={Photon-efficient quantum key distribution using time--energy entanglement with high-dimensional encoding},
  author={Zhong, Tian and Zhou, Hongchao and Horansky, Robert D and Lee, Catherine and Verma, Varun B and Lita, Adriana E and Restelli, Alessandro and Bienfang, Joshua C and Mirin, Richard P and Gerrits, Thomas and others},
  journal={New Journal of Physics},
  volume={17},
  number={2},
  pages={022002},
  year={2015},
doi={10.1088/1367-2630/17/2/022002},
  publisher={IOP Publishing}
}

@book{sakurai2020modern,
  title={Modern quantum mechanics},
  author={Sakurai, Jun John and Napolitano, Jim},
  year={2020},
  publisher={Cambridge University Press},
  url={https://www.cambridge.org/highereducation/books/modern-quantum-mechanics/DF43277E8AEDF83CC12EA62887C277DC#overview}
}

@article{benatti2020entanglement,
  title={Entanglement in indistinguishable particle systems},
  author={Benatti, Fabio and Floreanini, Roberto and Franchini, Fabio and Marzolino, Ugo},
  journal={Physics Reports},
  volume={878},
  pages={1--27},
  year={2020},
  publisher={Elsevier},
  doi={https://doi.org/10.1016/j.physrep.2020.07.003}
}

@article{xavier2025energy,
  title={Energy-time and time-bin entanglement: past, present and future},
  author={Xavier, Guilherme B and Larsson, Jan-{\AA}ke and Villoresi, Paolo and Vallone, Giuseppe and Cabello, Ad{\'a}n},
  journal={npj Quantum Information},
  volume={11},
  number={1},
  pages={129},
  year={2025},
  publisher={Nature Publishing Group UK London},
  doi={https://doi.org/10.1038/s41534-025-01072-3}
}

@article{montaut2025progress,
  title={Progress in integrated and fiber optics for time-bin based quantum information processing},
  author={Montaut, Nicola and George, Agnes and Monika, Monika and Nosrati, Farzam and Yu, Hao and Sciara, Stefania and Crockett, Benjamin and Peschel, Ulf and Wang, Zhiming and Lo Franco, Rosario and others},
  journal={Advanced Optical Technologies},
  volume={14},
  pages={1560084},
  year={2025},
  publisher={Frontiers Media SA},
  doi={https://doi.org/10.3389/aot.2025.1560084}
}

@article{reimer2019high,
  title={High-dimensional one-way quantum processing implemented on d-level cluster states},
  author={Reimer, Christian and Sciara, Stefania and Roztocki, Piotr and Islam, Mehedi and Romero Cort{\'e}s, Luis and Zhang, Yanbing and Fischer, Bennet and Loranger, S{\'e}bastien and Kashyap, Raman and Cino, Alfonso and others},
  journal={Nature Physics},
  volume={15},
  number={2},
  pages={148--153},
  year={2019},
  publisher={Nature Publishing Group UK London},
  doi={https://doi.org/10.1038/s41567-018-0347-x}
}

@article{brendel1999pulsed,
  title={Pulsed energy-time entangled twin-photon source for quantum communication},
  author={Brendel, J{\"u}rgen and Gisin, Nicolas and Tittel, Wolfgang and Zbinden, Hugo},
  journal={Physical Review Letters},
  volume={82},
  number={12},
  pages={2594},
  year={1999},
doi={https://doi.org/10.1103/PhysRevLett.82.2594},
  publisher={APS}
}

@article{cuevas2013long,
  title={Long-distance distribution of genuine energy-time entanglement},
  author={Cuevas, A and Carvacho, G and Saavedra, G and Cari{\~n}e, J and Nogueira, WAT and Figueroa, M and Cabello, Adan and Mataloni, P and Lima, G and Xavier, GB},
  journal={Nature Communications},
  volume={4},
  number={1},
  pages={2871},
  year={2013},
doi={10.1038/ncomms3871},
  publisher={Nature Publishing Group UK London}
}

@article{singh2025photonic,
      title={Photonic quantum information with time-bins: Principles and applications}, 
      author={Ashutosh Singh and Anuj Sethia and Leili Esmaeilifar and Raju Valivarthi and Neil Sinclair and Maria Spiropulu and Daniel Oblak},
      year={2025},
       journal={arXiv preprint arXiv:2507.08102},
      url={https://arxiv.org/abs/2507.08102}, 
}

@article{bouchard2020two,
  title={Two-photon interference: the Hong--Ou--Mandel effect},
  author={Bouchard, Fr{\'e}d{\'e}ric and Sit, Alicia and Zhang, Yingwen and Fickler, Robert and Miatto, Filippo M and Yao, Yuan and Sciarrino, Fabio and Karimi, Ebrahim},
  journal={Reports on Progress in Physics},
  volume={84},
  number={1},
  pages={012402},
  year={2020},
doi={10.1088/1361-6633/abcd7a},
  publisher={IOP Publishing}
}

@article{bergmayr2023harness,
   title={Harnessing high-dimensional temporal entanglement using limited interferometric setups},
   volume={22},
   ISSN={2331-7019},
   url={http://dx.doi.org/10.1103/PhysRevApplied.22.054054},
   DOI={10.1103/physrevapplied.22.054054},
   number={5},
   journal={Physical Review Applied},
   publisher={American Physical Society (APS)},
   author={Kanitschar, Florian and Bergmayr-Mann, Alexandra and Pivoluska, Matej and Huber, Marcus},
   year={2024},
   month=nov }

@article{Kanitschar_2025,
   title={Composable finite-size security of high-dimensional quantum-key-distribution protocols},
   volume={24},
   ISSN={2331-7019},
   url={http://dx.doi.org/10.1103/v51y-vkfr},
   number={5},
   journal={Physical Review Applied},
   publisher={American Physical Society (APS)},
   author={Kanitschar, Florian and Huber, Marcus},
   year={2025},
   month=nov }

@article{donohue2013coherent,
  title={Coherent ultrafast measurement of time-bin encoded photons},
  author={Donohue, John M and Agnew, Megan and Lavoie, Jonathan and Resch, Kevin J},
  journal={Physical Review Letters},
  volume={111},
  number={15},
  pages={153602},
  year={2013},
doi={https://doi.org/10.1103/PhysRevLett.111.153602},
  publisher={APS}
}

@inproceedings{Bennett_Brassard_1984,
  address = {India},
  author = {Bennett, C. H. and Brassard, G.},
  booktitle = {Proceedings of IEEE International Conference on Computers, Systems, and Signal Processing},
  location = {Bangalore},
  pages = 175,
  title = {{Quantum cryptography: Public key distribution and coin tossing}},
  year = 1984,
  publisher = {IEEE}
}

@article{Ekert_1991,
  title = {{Quantum cryptography based on Bell's theorem}},
  author = {Ekert, Artur K.},
  journal = {Phys. Rev. Lett.},
  volume = {67},
  issue = {6},
  pages = {661--663},
  numpages = {0},
  year = {1991},
  month = {Aug},
  publisher = {American Physical Society},
  doi = {10.1103/PhysRevLett.67.661},
  url = {https://link.aps.org/doi/10.1103/PhysRevLett.67.661}
}

@article{pe2005temporal,
  title={Temporal shaping of entangled photons},
  author={Pe'Er, Avi and Dayan, Barak and Friesem, Asher A and Silberberg, Yaron},
  journal={Physical Review Letters},
  volume={94},
  number={7},
  pages={073601},
  year={2005},
doi={https://doi.org/10.1103/PhysRevLett.94.073601},
  publisher={APS}
}

@article{tittel2000quantum,
  title={Quantum cryptography using entangled photons in energy-time Bell states},
  author={Tittel, Wolfgang and Brendel, J{\"u}rgen and Zbinden, Hugo and Gisin, Nicolas},
  journal={Physical Review Letters},
  volume={84},
  number={20},
  pages={4737},
  year={2000},
doi={https://doi.org/10.1103/PhysRevLett.84.4737},
  publisher={APS}
}

@article{tittel1998violation,
  title={Violation of Bell inequalities by photons more than 10 km apart},
  author={Tittel, Wolfgang and Brendel, J{\"u}rgen and Zbinden, Hugo and Gisin, Nicolas},
  journal={Physical Review Letters},
  volume={81},
  number={17},
  pages={3563},
  year={1998},
doi={https://doi.org/10.1103/PhysRevLett.81.3563},
  publisher={APS}
}

@article{franson1989bell,
  title={Bell inequality for position and time},
  author={Franson, James D},
  journal={Physical Review Letters},
  volume={62},
  number={19},
  pages={2205},
  year={1989},
doi={https://doi.org/10.1103/PhysRevLett.62.2205},
  publisher={APS}
}

@article{sun2016quantum,
  title={Quantum teleportation with independent sources and prior entanglement distribution over a network},
  author={Sun, Qi-Chao and Mao, Ya-Li and Chen, Si-Jing and Zhang, Wei and Jiang, Yang-Fan and Zhang, Yan-Bao and Zhang, Wei-Jun and Miki, Shigehito and Yamashita, Taro and Terai, Hirotaka and others},
  journal={Nature Photonics},
  volume={10},
  number={10},
  pages={671--675},
  year={2016},
doi={http://dx.doi.org/10.1038/nphoton.2016.180},
  publisher={Nature Publishing Group UK London}
}

@article{Freq1,
doi = {10.1088/1367-2630/11/10/103052},
url = {https://dx.doi.org/10.1088/1367-2630/11/10/103052},
year = {2009},
month = {oct},
publisher = {},
volume = {11},
number = {10},
pages = {103052},
author = {Fedrizzi, Alessandro and Herbst, Thomas and Aspelmeyer, Markus and Barbieri, Marco and Jennewein, Thomas and Zeilinger, Anton},
title = {Anti-symmetrization reveals hidden entanglement},
journal = {New Journal of Physics}
}

@article{Freq2,
author = {S. Francesconi and F. Baboux and A. Raymond and N. Fabre and G. Boucher and A. Lema\^{i}tre and P. Milman and M. I. Amanti and S. Ducci},
journal = {Optica},
number = {4},
pages = {316--322},
publisher = {Optica Publishing Group},
title = {Engineering two-photon wavefunction and exchange statistics in a semiconductor chip},
volume = {7},
month = {Apr},
year = {2020},
url = {https://opg.optica.org/optica/abstract.cfm?URI=optica-7-4-316},
doi = {10.1364/OPTICA.379477},
}

@article{TransverseSM,
  title = {Generation of a Two-Photon Singlet Beam},
  author = {Nogueira, W. A. T. and Walborn, S. P. and P\'adua, S. and Monken, C. H.},
  journal = {Phys. Rev. Lett.},
  volume = {92},
  issue = {4},
  pages = {043602},
  numpages = {4},
  year = {2004},
  month = {Jan},
  publisher = {American Physical Society},
  doi = {10.1103/PhysRevLett.92.043602},
  url = {https://link.aps.org/doi/10.1103/PhysRevLett.92.043602}
}

@article{fabre2022hong,
  title={The Hong--Ou--Mandel experiment: from photon indistinguishability to continuous-variable quantum computing},
  author={Fabre, Nicolas and Amanti, Maria and Baboux, Florent and Keller, Arne and Ducci, Sara and Milman, P{\'e}rola},
  journal={The European Physical Journal D},
  volume={76},
  number={10},
  pages={196},
  year={2022},
  publisher={Springer},
  doi={https://doi.org/10.1140/epjd/s10053-022-00525-0}
}

@article{
OAM1,
author = {Yingwen Zhang  and Filippus S. Roux  and Thomas Konrad  and Megan Agnew  and Jonathan Leach  and Andrew Forbes },
title = {Engineering two-photon high-dimensional states through quantum interference},
journal = {Science Advances},
volume = {2},
number = {2},
pages = {e1501165},
year = {2016},
doi = {10.1126/sciadv.1501165}}

@article{OAM2,
  title = {Imaging symmetric and antisymmetric behavior of orbital-angular-momentum-entangled two-photon states},
  author = {Ibarra-Borja, Zeferino and Yepiz-Graciano, Pablo and Claro-Rodr\'{\i}guez, Nicolas and U'Ren, Alfred B. and Ram\'{\i}rez-Alarc\'on, Roberto},
  journal = {Phys. Rev. Appl.},
  volume = {22},
  issue = {2},
  pages = {024068},
  numpages = {12},
  year = {2024},
  month = {Aug},
  publisher = {American Physical Society},
  doi = {10.1103/PhysRevApplied.22.024068},
  url = {https://link.aps.org/doi/10.1103/PhysRevApplied.22.024068}
}

@article{OAM3,
  title = {Hong-Ou-Mandel Interference between Two Hyperentangled Photons Enables Observation of Symmetric and Antisymmetric Particle Exchange Phases},
  author = {Liu, Zhi-Feng and Chen, Chao and Xu, Jia-Min and Cheng, Zi-Mo and Ren, Zhi-Cheng and Dong, Bo-Wen and Lou, Yan-Chao and Yang, Yu-Xiang and Xue, Shu-Tian and Liu, Zhi-Hong and Zhu, Wen-Zheng and Wang, Xi-Lin and Wang, Hui-Tian},
  journal = {Phys. Rev. Lett.},
  volume = {129},
  issue = {26},
  pages = {263602},
  numpages = {7},
  year = {2022},
  month = {Dec},
  publisher = {American Physical Society},
  doi = {10.1103/PhysRevLett.129.263602},
  url = {https://link.aps.org/doi/10.1103/PhysRevLett.129.263602}
}

@article{Terhal_2000,
  title = {Schmidt number for density matrices},
  author = {Terhal, Barbara M. and Horodecki, Pawe\l{}},
  journal = {Phys. Rev. A},
  volume = {61},
  issue = {4},
  pages = {040301},
  numpages = {4},
  year = {2000},
  month = {Mar},
  publisher = {American Physical Society},
  doi = {10.1103/PhysRevA.61.040301},
  url = {https://link.aps.org/doi/10.1103/PhysRevA.61.040301}
}

@inproceedings{Lofberg2004,
address = {Taipei, Taiwan},
author = {L{\"{o}}fberg, J.},
booktitle = {In Proceedings of the CACSD Conference},
title = {YALMIP : A Toolbox for Modeling and Optimization in MATLAB},
year = {2004}
}

@manual{mosek,
author = "MOSEK ApS",
title = "The MOSEK optimization toolbox for MATLAB manual. Version 9.0.",
year = 2019,
url = "http://docs.mosek.com/9.0/toolbox/index.html"}

@article{esmaeil2020efficient,
  title={Efficient single-photon detection with 7.7 ps time resolution for photon-correlation measurements},
  author={Esmaeil Zadeh, Iman and Los, Johannes WN and Gourgues, Ronan BM and Chang, Jin and Elshaari, Ali W and Zichi, Julien Romain and Van Staaden, Yuri J and Swens, Jeroen PE and Kalhor, Nima and Guardiani, Antonio and others},
  journal={Acs Photonics},
  volume={7},
  number={7},
  pages={1780--1787},
  year={2020},
  publisher={ACS Publications},
  url={https://doi.org/10.1021/acsphotonics.0c00433}
}

@article{korzh2020demonstration,
  title={Demonstration of sub-3 ps temporal resolution with a superconducting nanowire single-photon detector},
  author={Korzh, Boris and Zhao, Qing-Yuan and Allmaras, Jason P and Frasca, Simone and Autry, Travis M and Bersin, Eric A and Beyer, Andrew D and Briggs, Ryan M and Bumble, Bruce and Colangelo, Marco and others},
  journal={Nature Photonics},
  volume={14},
  number={4},
  pages={250--255},
  year={2020},
  publisher={Nature Publishing Group UK London},
  url={https://doi.org/10.1038/s41566-020-0589-x}
}

@article{Chang2021,
author={Chang, Kai-Chi
and Cheng, Xiang
and Sarihan, Murat Can
and Vinod, Abhinav Kumar
and Lee, Yoo Seung
and Zhong, Tian
and Gong, Yan-Xiao
and Xie, Zhenda
and Shapiro, Jeffrey H.
and Wong, Franco N. C.
and Wong, Chee Wei},
title={648 Hilbert-space dimensionality in a biphoton frequency comb: entanglement of formation and Schmidt mode decomposition},
journal={npj Quantum Information},
year={2021},
month={Mar},
day={11},
volume={7},
number={1},
pages={48},
issn={2056-6387},
doi={10.1038/s41534-021-00388-0},
url={https://doi.org/10.1038/s41534-021-00388-0}
}

@article{HerreraValencia2020highdimensional,
  doi = {10.22331/q-2020-12-24-376},
  url = {https://doi.org/10.22331/q-2020-12-24-376},
  title = {High-{D}imensional {P}ixel {E}ntanglement: {E}fficient {G}eneration and {C}ertification},
  author = {Herrera Valencia, Natalia and Srivastav, Vatshal and Pivoluska, Matej and Huber, Marcus and Friis, Nicolai and McCutcheon, Will and Malik, Mehul},
  journal = {{Quantum}},
  issn = {2521-327X},
  publisher = {{Verein zur F{\"{o}}rderung des Open Access Publizierens in den Quantenwissenschaften}},
  volume = {4},
  pages = {376},
  month = dec,
  year = {2020}
}

@article{Cheng2023,
author={Cheng, Xiang
and Chang, Kai-Chi
and Sarihan, Murat Can
and Mueller, Andrew
and Spiropulu, Maria
and Shaw, Matthew D.
and Korzh, Boris
and Faraon, Andrei
and Wong, Franco N. C.
and Shapiro, Jeffrey H.
and Wong, Chee Wei},
title={High-dimensional time-frequency entanglement in a singly-filtered biphoton frequency comb},
journal={Communications Physics},
year={2023},
month={Sep},
day={28},
volume={6},
number={1},
pages={278},
issn={2399-3650},
doi={10.1038/s42005-023-01370-2},
url={https://doi.org/10.1038/s42005-023-01370-2}
}

@article{Ndagano2020,
author={Ndagano, Bienvenu
and Defienne, Hugo
and Lyons, Ashley
and Starshynov, Ilya
and Villa, Federica
and Tisa, Simone
and Faccio, Daniele},
title={Imaging and certifying high-dimensional entanglement with a single-photon avalanche diode camera},
journal={npj Quantum Information},
year={2020},
month={Dec},
day={04},
volume={6},
number={1},
pages={94},
issn={2056-6387},
doi={10.1038/s41534-020-00324-8},
url={https://doi.org/10.1038/s41534-020-00324-8}
}

@article{PRXQuantum.4.010308,
  title = {Manipulation and Certification of High-Dimensional Entanglement through a Scattering Medium},
  author = {Courme, Baptiste and Cameron, Patrick and Faccio, Daniele and Gigan, Sylvain and Defienne, Hugo},
  journal = {PRX Quantum},
  volume = {4},
  issue = {1},
  pages = {010308},
  numpages = {15},
  year = {2023},
  month = {Jan},
  publisher = {American Physical Society},
  doi = {10.1103/PRXQuantum.4.010308},
  url = {https://link.aps.org/doi/10.1103/PRXQuantum.4.010308}
}

@article{Nape2021,
author={Nape, Isaac
and Rodr{\'i}guez-Fajardo, Valeria
and Zhu, Feng
and Huang, Hsiao-Chih
and Leach, Jonathan
and Forbes, Andrew},
title={Measuring dimensionality and purity of high-dimensional entangled states},
journal={Nature Communications},
year={2021},
month={Aug},
day={27},
volume={12},
number={1},
pages={5159},
issn={2041-1723},
doi={10.1038/s41467-021-25447-0},
url={https://doi.org/10.1038/s41467-021-25447-0}
}

@article{
MarioKrenn2014,
author = {Mario Krenn  and Marcus Huber  and Robert Fickler  and Radek Lapkiewicz  and Sven Ramelow  and Anton Zeilinger },
title = {Generation and confirmation of a (100 × 100)-dimensional entangled quantum system},
journal = {Proceedings of the National Academy of Sciences},
volume = {111},
number = {17},
pages = {6243-6247},
year = {2014},
URL = {https://www.pnas.org/doi/abs/10.1073/pnas.1402365111}}

@article{Bavaresco2018,
author={Bavaresco, Jessica
and Herrera Valencia, Natalia
and Kl{\"o}ckl, Claude
and Pivoluska, Matej
and Erker, Paul
and Friis, Nicolai
and Malik, Mehul
and Huber, Marcus},
title={Measurements in two bases are sufficient for certifying high-dimensional entanglement},
journal={Nature Physics},
year={2018},
month={Oct},
day={01},
volume={14},
number={10},
pages={1032-1037},
issn={1745-2481},
doi={10.1038/s41567-018-0203-z},
url={https://doi.org/10.1038/s41567-018-0203-z}
}

@article{schiffer2026brightsourcehighdimensionaltemporal,
      title={Bright Source of High-Dimensional Temporal Entanglement}, 
      author={Dorian Schiffer and Robert Kindler and Alexandra Bergmayr-Mann and Florian Kanitschar and Amin Babazadeh and Paul Erker and Marcus Huber and Anton Zeilinger},
      year={2026},
      journal={arXiv preprint arXiv:2601.07678},
      primaryClass={quant-ph},
      url={https://arxiv.org/abs/2601.07678}, 
}

@article{ponce2023,
author = {Cabrejo-Ponce, Meritxell and Muniz, André Luiz Marques and Huber, Marcus and Steinlechner, Fabian},
title = {High-Dimensional Entanglement for Quantum Communication in the Frequency Domain},
journal = {Laser \& Photonics Reviews},
volume = {17},
number = {9},
pages = {2201010},
url = {https://onlinelibrary.wiley.com/doi/abs/10.1002/lpor.202201010},
year = {2023}
}

@article{xiaomin2020,
  title = {Efficient Generation of High-Dimensional Entanglement through Multipath Down-Conversion},
  author = {Hu, Xiao-Min and Xing, Wen-Bo and Liu, Bi-Heng and Huang, Yun-Feng and Li, Chuan-Feng and Guo, Guang-Can and Erker, Paul and Huber, Marcus},
  journal = {Phys. Rev. Lett.},
  volume = {125},
  issue = {9},
  pages = {090503},
  numpages = {6},
  year = {2020},
  month = {Aug},
  publisher = {American Physical Society},
  doi = {10.1103/PhysRevLett.125.090503},
  url = {https://link.aps.org/doi/10.1103/PhysRevLett.125.090503}
}

@article{Courme2023,
author = {Baptiste Courme and Chlo\'{e} Verni\`{e}re and Peter Svihra and Sylvain Gigan and Andrei Nomerotski and Hugo Defienne},
journal = {Opt. Lett.},
number = {13},
pages = {3439--3442},
publisher = {Optica Publishing Group},
title = {Quantifying high-dimensional spatial entanglement with a single-photon-sensitive time-stamping camera},
volume = {48},
month = {Jul},
year = {2023},
url = {https://opg.optica.org/ol/abstract.cfm?URI=ol-48-13-3439},
doi = {10.1364/OL.487182},
}

\appendix
\onecolumngrid

\section{Measurements Applied to Maximally Entangled States} \label{apdx:measurementtheory}
In this section, we aim to gain an intuition for the experimental setup that allows us to perform two distinct transformations. First, we may insert the Mach-Zehnder interferometer and apply a negative phase shift in the long arm. Second, we can insert the MZI and apply a positive phase shift instead. To understand the effect of these operations, let us, for now, consider a source that prepares the maximally entangled state $\ket{\Phi^+} = \frac{1}{\sqrt{d}} \sum_{i} \ket{ii}$. 

For the first case, where we insert an MZI and apply a negative phase shift, we obtain

\begin{align*}
    \ket{\Phi^+} &\stackrel{\text{MZI}}{\rightarrow} \frac{1}{\sqrt{2d}} \sum_{i=0}^{d-1} \left(\ket{i+2} -\ket{i} \right)\ket{i}\\
    & \stackrel{\text{delay}}{\rightarrow}  \frac{1}{\sqrt{2d}} \sum_{i=0}^{d-1} \left(\ket{i+2} -\ket{i} \right)\ket{i+1}\\
    &~~~~=: \ket{\Phi_1}.
\end{align*}

The (pseudo-) unitary (up to boundary effects) representing this setup reads $U := \frac{\hat{T}_{+2} - \mathbbm{1}}{2} \otimes \hat{T}_{+1}$, where $\hat{T}$ represents the time-shift operator defined by $\hat{T}_{+t} \ket{T} = \ket{T+t}$.

Similarly, for the case with the MZI and a positive phase, we obtain

\begin{align*}
    \ket{\Phi^+} &\stackrel{\text{MZI}}{\rightarrow} \frac{1}{\sqrt{2d}} \sum_{i=0}^{d-1} \left(\ket{i+2} +\ket{i} \right)\ket{i}\\
    & \stackrel{\text{delay}}{\rightarrow}  \frac{1}{\sqrt{2d}} \sum_{i=0}^{d-1} \left(\ket{i+2} +\ket{i} \right)\ket{i+1}\\
    &~~~~=: \ket{\Phi_2}.
\end{align*}

Our goal is to certify the dimension of the generated state while only being able to count the clicks at the output of the setup (i.e., without being able to resolve the dimension). 

Define the swap operator $S:= \sum_{i,j} \ket{ij}\!\!\bra{ji}$. By definition, $\ket{\Phi^+}$ is symmetric under swapping,

\begin{align*}
    S\ket{\Phi^+} &= \frac{1}{\sqrt{d}} \sum_{i,j} \ket{ij}\!\!\bra{ji} \sum_{k=0}^{d-1} \ket{kk} = \frac{1}{\sqrt{d}} \sum_{i,j,k} \delta_{i,k} \delta_{j,k} \ket{ij}\\
    &= \frac{1}{\sqrt{d}} \sum_{i=0}^{d-1} \ket{ii}.
\end{align*}

In contrast, $\ket{\Phi_1}$ turns out to be partially (the amount of it depends on the dimension) anti-symmetric under swap,

\begin{align*}
    S\ket{\Phi_1} &= \sum_{i,j=0}^{d-1} \ket{ij}\!\!\bra{ji} \frac{1}{\sqrt{2d}} \sum_{k=0}^{d-1} \left(\ket{k+2} -\ket{k} \right)\ket{k+1}\\
    &=\frac{1}{\sqrt{2d}} \sum_{i,j, k=0}^{d-1} \left(\delta_{j,k+2} \delta_{i,k+1} - \delta_{j,k} \delta_{i,k+1}\right) \ket{ij}\\
    &= \frac{1}{\sqrt{2d}}\left( \sum_{k=0}^{d-1} \ket{k+1,k+2} - \sum_{k=0}^{d-1} \ket{k+1,k}  \right) \\
    &= -\frac{1}{\sqrt{2d}}\left( \sum_{k=0}^{d-1} \ket{k+1,k} -  \sum_{k=0}^{d-1} \ket{k+1,k+2}  \right) \\
    &= -\frac{1}{\sqrt{2d}}\left( \sum_{i=-1}^{d-2} \ket{i+2,i+1} -  \sum_{i=1}^{d} \ket{i,i+1}  \right)\\
    &\approx -\frac{1}{\sqrt{2d}}\left( \sum_{i=0}^{d-2} \ket{i+2,i+1} -  \sum_{i=0}^{d-1} \ket{i,i+1}  \right)\\
    &\approx -\ket{\Phi_1}.
\end{align*}

Similarly, we obtain

\begin{align*}
    S\ket{\Phi_2} \approx \ket{\Phi_2}.
\end{align*}

Summing up, $\ket{\Phi_1}$ is partially  anti-symmetric under swap, while $\ket{\Phi_2}$  and $\ket{\Phi^+}$ are, respectively, partially and fully  symmetric. We aim to use this information to certify entanglement and dimensionality of entanglement. Note that we first discretized the continuous space into a $d$-dimensional time-bin space. Consequently, the applied delay pushes parts of the state outside of the original $d$-dimensional space, affecting the observed probabilities when measurements are applied. In this work, we deliberately do not extend the Hilbert space used to describe the experiment, but project back onto the original $d$-dimensional space, which carries the implicit assumption that the delay acts equivalently on all time bins, which is justified and reasonable in our considered scenario.

Finally, define the projectors onto the symmetric and anti-symmetric subspaces,

\begin{align}
    \Pi_S &:= \frac{1}{2} \left( \mathbbm{1}+ S \right)\\
    \Pi_A &:= \frac{1}{2} \left( \mathbbm{1}- S \right),
\end{align}

and note that (in the limit of $d\rightarrow \infty$)

\begin{align*}
    \Tr{\Pi_S \ketbra{\Phi_{2}}} &= \Tr{\frac{1}{2}\left( \mathbbm{1}+S\right) \ketbra{\Phi_{2}}}= \frac{1}{2} \left( 1+ \Tr{\ketbra{\Phi_{2}}} \right) = 1\\
    \Tr{\Pi_A \ketbra{\Phi_{2}}} &= \Tr{\frac{1}{2}\left( \mathbbm{1}-S\right) \ketbra{\Phi_{2}}}= \frac{1}{2} \left( 1- \Tr{\ketbra{\Phi_{2}}} \right) = 0\\
    \Tr{\Pi_S \ketbra{\Phi_1}} &= \Tr{\frac{1}{2}\left( \mathbbm{1}+S\right) \ketbra{\Phi_1}} = \frac{1}{2} \left( 1- \Tr{\ketbra{\Phi_1}} \right) = 0\\
    \Tr{\Pi_A \ketbra{\Phi_1}} &= \Tr{\frac{1}{2}\left( \mathbbm{1}-S\right) \ketbra{\Phi_1}} = \frac{1}{2} \left( 1+ \Tr{\ketbra{\Phi_1}} \right) = 1.
\end{align*}

\section{Proof of the Schmidt-Number witness} \label{apdx:derivationWitness}
Our goal is to certify the Schmidt number of the unknown source state. It is known that the fidelity $F$ of a state $\rho$ with a maximally entangled state of dimension $d$ is related to the Schmidt number $k$ of $\rho$ via the inequality $F \leq \frac{k}{d}$ \cite{Terhal_2000}. One could therefore attempt to certify entanglement by directly lower-bounding the fidelity of the unknown source state with respect to a suitably chosen target state. Although this approach performs well in low dimensions, it yields poor fidelity certificates already in the low double-digit regime. This behavior can be understood as follows. To obtain tight bounds while retaining full generality, we use a semidefinite program (see Appendix \ref{apdx:SDP_Derivation}) to compute lower bounds on the fidelity from our measurement data. For practical reasons, we require that the number of measurement settings remain fixed rather than scaling with the dimension. As the dimension increases, the set of states compatible with this fixed amount of data grows. Because the SDP optimizes over all states consistent with the measurements and returns the worst-case one, the resulting fidelity bounds decrease with dimension, leading in turn to smaller certified Schmidt numbers.

We circumvent this problem by proposing and proving a new Schmidt-number witness that enables significantly higher certified Schmidt numbers. Intuitively, instead of comparing the unknown state to a single target state, we consider an entire family of target states and evaluate how close the unknown source state is to the best-fitting member of this family. More precisely, we show that for every state $\rho \in \mathcal{D}(\mathcal{H}_1 \otimes \mathcal{H}_2)$ and for the witness $W(\rho) = \frac{1}{d} \sum_{i,j=0}^{d-1} |\langle ii|\rho|jj\rangle|$ defined with respect to orthonormal bases $\{\ket{i}\}_{i=0}^{d-1}$ and $\{\ket{j}\}_{j=0}^{d-1}$ for $\mathcal{H}_1$ and $\mathcal{H}_2$, respectively, the following relation holds

\begin{equation}
    \max_{\rho \in \mathcal{S}_k} W(\rho) \leq k / d \;,
\end{equation}

for $\mathcal{S}_k$, the subset of all density matrices with Schmidt number at most $k$. In turn, observing $W(\rho) > k/d$, implies $\rho \notin \mathcal{S}_k$, i.e., $\rho$ has at least Schmidt-number $k+1$. 

Due to convexity we can reduce the maximum over all states with Schmidt number of $k$ or lower by a maximum over pure, bipartite states of Schmidt rank at most $k$, $\ket{\Psi_k} = \sum_{m=0}^{k-1} \lambda_m \ket{m_A m_B}$, where $\{\ket{m_A}\}_{m=0}^{k-1}$ and $\{\ket{m_B}\}_{m=0}^{k-1}$ are the Schmidt bases of the state $\ket{\Psi_k}$. This leads to

\begin{align*}
   & \max_{\rho \in \mathcal{S}_k} W(\rho) = \max_{\ket{\Psi_k}} W(\ketbra{\Psi_k}) = \frac{1}{d} \max_{\lambda} \sum_{i,j=0}^{d-1} \left| \bra{ii} \left(\sum_{m=0}^{k-1} \lambda_m \ket{m_A m_B} \sum_{n=0}^{k-1} \lambda_n \bra{n_A n_B} \right) \ket{jj} \right| \\
   &\stackrel{\Delta \text{-inequ.}}{\leq} \frac{1}{d} \max_{\lambda} \sum_{m,n=0}^{k-1} \lambda_m \lambda_n \left(\sum_{i=0}^{d-1} \left| \bra{ii}m_A m_B\rangle\right|\right) \left(\sum_{j=0}^{d-1} \left|\bra{n_A n_B} jj\rangle \right|\right)\;,
\end{align*}

where we used the fact that the Schmidt coefficients are positive, real numbers. Note that we can expand the bases $\{\ket{i}\}_{i=0}^{d-1}$ and $\{\ket{j}\}_{j=0}^{d-1}$ in terms of the Schmidt bases, $\ket{i} = \sum_m a_m^i \ket{m_A}$ and $\ket{j} = \sum_n b_n^j \ket{m_A}$ with $\sum_{i=0}^{d-1} |a_m^i|^2 = 1 = \sum_{j=0}^{d-1} |b_n^j|^2 $. This allows us to bound the inner sums as follows,

\begin{align}
    &\sum_{i=0}^{d-1} \left| \bra{ii}m_A m_B\rangle\right| = \sum_{i=0}^{d-1} \left|a_m^i b_m ^i\right| = \sum_{i=0}^{d-1} \left|a_m^i\right| \left| b_m ^i\right| \leq \sqrt{\sum_{i=0}^{d-1} \left|a_m^i\right|^2} \sqrt{\sum_{i=0}^{d-1} \left|b_m^i\right|^2} \leq 1,\\
    &\sum_{j=0}^{d-1} \left| \bra{n_A n_B}jj\rangle\right| = \sum_{i=0}^{d-1} \left|a_n^i b_n ^i\right| = \sum_{i=0}^{d-1} \left|a_n^i\right| \left| b_n ^i\right| \leq \sqrt{\sum_{i=0}^{d-1} \left|a_n^i\right|^2} \sqrt{\sum_{i=0}^{d-1} \left|b_n^i\right|^2} \leq 1\;,
\end{align}

where we used the Cauchy-Schwarz inequality in the second last step. Thus, we obtain

\begin{align}
     \frac{1}{d} \max_{\lambda} \sum_{m,n=0}^{k-1} \lambda_m \lambda_n \left(\sum_{i=0}^{d-1} \left| \bra{ii}m_A m_B\rangle\right|\right) \left(\sum_{j=0}^{d-1} \left|\bra{n_A n_B} jj\rangle \right|\right) \leq  \frac{1}{d} \max_{\lambda} \sum_{m,n=0}^{k-1} \lambda_m \lambda_n =  \frac{1}{d} \max_{\lambda} \left(\sum_{n=0}^{k-1}  \lambda_n\right)^2\;.
\end{align}

Now, observe that

\begin{align}
\left(\sum_{n=0}^{k-1} \lambda_n\right)^2 = \left| \sum_{n=0}^{k-1} \lambda_n \cdot 1 \right|^2 \leq \sum_{n=0}^{k-1} \lambda_n^2 \sum_{m=0}^{k-1} 1^2 = 1 \cdot k\;,
\end{align}

where we used that for Schmidt-coefficients, we have $\sum_{n=0}^{k-1} \lambda_n ^2 = 1$. Putting things together, we obtain

\begin{equation}
    \max_{\rho \in \mathcal{S}_k} W(\rho) \leq \frac{k}{d}\;.
\end{equation}

\section{Derivation of the optimization problem} \label{apdx:SDP_Derivation}
Our goal is to derive a lower bound on the Schmidt number, based on the Schmidt number witness derived in the previous section, $W(\rho) =\frac{1}{d} \sum_{i,j} |\bra{ii}\rho\ket{jj}|$.  Therefore, we use the experimental constraints from Appendix \ref{apdx:measurementtheory}. Let $U_+$ and $U_-$ denote the transformations representing the action of the setup on the source state, respectively. Then, the constraints due to our experimental observations take the form

\begin{align*}
\Tr{\rho_- \Pi_A} &= \Tr{U_- \rho U_-^{\dagger} \Pi_A}  = \Tr{U_- \rho U_-^{\dagger} \frac{1}{2}\left( \mathbbm{1}-S \right)} = \frac{1}{2} - \frac{1}{2} \Tr{U_-\rho U_-^{\dagger} \sum_{i,j}\ket{i,j}\!\!\bra{j,i}}\\
&= \frac{1}{2} - \frac{1}{2} \sum_{i,j} \bra{j,i} U_-\rho U_-^{\dagger}\ket{i,j} = \frac{CC_{\textrm{HOM}}^{\textrm{-}}}{N^{\textrm{-}}}=: C_{\text{exp, -}}\;,    
\end{align*}
and similarly,
\begin{align*}
\Tr{\rho_+ \Pi_A} &= \frac{1}{2} - \frac{1}{2} \sum_{i,j} \bra{j,i} U_+\rho U_+^{\dagger}\ket{i,j}= \frac{CC_{\textrm{HOM}}^{\textrm{+}}}{N^{\textrm{+}}}=: C_{\text{exp, +}},   
\end{align*}

where $CC_{\textrm{HOM}}^{\pm}$ are the observed coincidence counts in the HOM measurement after the transformations $U_{\pm}$, and $N^{\pm}$ are the corresponding total photon pairs entering the HOM measurement. For the first two conditions, we use the cyclic property of the trace to move the transformation over to the operator and away from the state,

\begin{align}
    U_\pm \ket{i,j} &= \frac{1}{\sqrt{2}} \left( \ket{i} \pm \ket{i+2} \right) \ket{j+1}\; .
\end{align}
To ease notation, define
\begin{align}\label{eq:def:S_m}
    S_\pm := U_\pm^{\dagger} S U_\pm \;.
\end{align}

This allows us to rewrite the first two constraints as

\begin{align*}
    &\Tr{\rho S_\pm}= \frac{1}{2}\sum_{i,j} \left[ \left( \bra{j+2} \pm \bra{j} \right) \bra{i+1}\right] \rho \left[\left(\ket{i+1} \pm \ket{i} \right)\ket{j+1}\right]\\
    &= \frac{1}{2}\sum_{i,j} \left( \bra{j,i+1}\rho\ket{i,j+1}\pm \bra{j,i+1}\rho\ket{i+2,j+1} \right.\\
    &~~~~~~~~\left. \pm \bra{j+2, i+1}\rho\ket{i,j+1} + \bra{j+2, i+1}\rho\ket{i+2, j+1} \right),
\end{align*}

which leads to the rewritten forms

\begin{align*}
   & \Tr{\rho S_+} = 1-2 C_{\textrm{exp, +}}\\
   & \Tr{\rho S_-} = 1-2 C_{\textrm{exp, -}}\;\, ,
\end{align*}

As we aim for the state $\rho$ which minimizes the Fidelity to the class of entangled states $\Phi^+_n$ (defined in Sec. \ref{TheoryMethod} and containing $\Phi^+$, that is, our suspected target state) while satisfying all those constraints, we obtain

\begin{align*}
    \mathrm{minimise  }&~~ \frac{1}{d} \sum_{m,n=0}^{d-1} | \bra{mm}\rho\ket{nn}|\\
    \mathrm{s.t.: }&\\
    & \Tr{\rho S_+} = 1 - 2 C_{\text{exp, +}}\\
    & \Tr{\rho S_-} = 1 - 2 C_{\text{exp, -}}\\
    &\Tr{\rho} = 1\\
    & \rho \geq 0.
\end{align*}

As we cannot expect arbitrary measurement accuracy, we relax those constraints to inequalities. Let us denote lower and upper bounds (retrieved from the experimental values and errors) on our experimental observations by superscripts $\ell$ and $u$. Then, the relaxed problem reads

\begin{equation}\label{eq:PrimalProblem}
  \begin{aligned}
    \mathrm{minimize  }&~~ \frac{1}{d} \sum_{m,n=0}^{d-1} | \bra{mm}\rho\ket{nn}|\\
    \mathrm{s.t.: }&\\
    & 1- 2 C_{\text{exp, +}}^{u} \leq \Tr{\rho S_+} \leq 1- 2 C_{\text{exp, +}}^{\ell}\\
    & 1- 2 C_{\text{exp, -}}^{u} \leq \Tr{\rho S_-} \leq 1- 2 C_{\text{exp, -}}^{\ell}\\
    &\Tr{\rho} = 1\\
    & \rho \geq 0.
\end{aligned}  
\end{equation}

This SDP represents the core problem based on experimental observations. However, we may want to include physically motivated assumptions, which are discussed in the following sections.

\section{Time invariance}\label{apdx:TimeInv}
In the considered setup, the laser used to pump the SPDC crystal operates in the continuous-wave regime, such that the temporal distribution of SPDC events is effectively uniform in time. This implies a certain symmetry between subblocks of the density matrix of the unknown quantum state. In more detail, the $(d+2)\times(d+2)$ matrices with corners in the matrix elements $\bra{ii}\rho\ket{ii}$, $\bra{ii}\rho\ket{i+1,i+1}$, $\bra{i+1,i+1}\rho\ket{ii}$ and $\bra{i+1,i+1}\rho\ket{i+1,i+1}$ must be equal for all $i \in\{0,...,d-2\}$, which implies that $(d+2) \times (d+2)$- diagonal submatrices of the state's density matrix are equal. We can incorporate this symmetry as an additional constraint in our optimization problem.

Define the sets $I_j := \{j \cdot d, j \cdot d+1,...  d+j\cdot d\}$, indexed by $j$ and denote by $\rho(I_j, I_j) = \rho(j \cdot d : d+j\cdot d, j\cdot d : d+j\times d$. Then this boils down to the additional constraint $\rho(I_j, I_j) =  \rho(I_0, I_0)$.

While one could express this directly as constraints for the SDP, using block-wise equality, e.g., $\rho(1:d+1, 1:d+1) = \rho(n(d+1):(n+1)(d+1))$, it turns out that it is more convenient to formulate this condition as a linear map. Specifically, $\mathcal{T}_n(\rho) := T_n^{\dagger} \rho T_n - T_0 \rho T_0^{\dagger}$, where $T_n$, $n \in \{1,..., d-1\}$ are $d^2 \times (d+2)$-dimensional matrices, used to extract $(d+2) \times (d+2)$-dimensional diagonal blocks from $\rho$,

\begin{align}
    T_0 := \begin{pmatrix}
        \mathbbm{1}_{(d+2) \times (d+2)}\\
        0_{(d+1) \times (d+1)} \\
        0_{(d+1) \times (d+1)} \\
        \vdots \\
        0_{(d+1) \times (d+1)}
    \end{pmatrix}, ~~~ T_1 = \begin{pmatrix}
        0_{(d+1) \times (d+1)} \\
        \mathbbm{1}_{(d+2) \times (d+2)}\\
        0_{(d+1) \times (d+1)} \\
        \vdots \\
        0_{(d+1) \times (d+1)}
    \end{pmatrix}, ...\, , T_{d-1} = \begin{pmatrix}
        0_{(d+1) \times (d+1)} \\
        0_{(d+1) \times (d+1)} \\
        \vdots \\
        \mathbbm{1}_{(d+2) \times (d+2)}
    \end{pmatrix}. 
\end{align}

This adds the constraints

\begin{align*}
    \mathcal{T}_n(\rho) = T_n^{\dagger} \rho T_n - T_0 \rho T_0^{\dagger} = 0 ~~~ \forall n \in \{1,...,d-1\}
\end{align*}

to our primal optimization problem. This reformulation of the symmetry condition into constraints using linear maps eases the transformation of the whole primal SDP into its dual form later on.

\section{Diagonal symmetry}\label{apdx:SymmDiag}
Motivated by the independence of the pair production process from exterior noises, such as dark counts eventually creating pairs of photons at different times $i \ne j$, we observe another symmetry in the unknown source state. Specifically, this implies a symmetry in the state's diagonal elements, $\forall i \in \{0,..., d-1\}:~ \bra{ii}\rho\ket{ii} = \bra{00}\rho\ket{00}$ and $\forall i\neq j:~\bra{ij}\rho\ket{ij} = \bra{01}\rho\ket{01}$. As before, we prefer to express this symmetry condition in the form of linear maps. Therefore, let us introduce the maps $Q_{i,j} = \ket{i,j}\!\!\bra{i,j}$, which leads to the constraints 

\begin{align*}
    \mathcal{Q}^1_{i}(\rho) &= \Tr{\rho Q_{i,i}} - \Tr{\rho Q_{0,0}} = 0~~~\forall i \in \{1, ..., d-1 \}\\
    \mathcal{Q}^2_{i,j}(\rho) &= \Tr{\rho Q_{i,j}} - \Tr{\rho Q_{0,1}} = 0 ~~~\forall i\neq j \in \{0,...,d-1\}.
\end{align*}

Having formulated both symmetries as linear maps, we can incorporate those constraints into the primal optimization problem given in Eq. (\ref{eq:PrimalProblem}).

\section{Transformation to the Dual}\label{apdx:TransformationDual}
While the current formulation of the semidefinite program is physically intuitive and conceptually clear, it presents two significant disadvantages. First, the size of the optimization variable $\rho$ scales quadratically with the Hilbert space dimension, which makes the problem computationally expensive even for moderate, low-double-digit dimensions. Second, since we solve the SDP numerically, we cannot expect to attain the minimum with arbitrary precision. As a result, we typically obtain only upper bounds on the witness value, rather than tight exact values.

To simplify notation, let us introduce

\begin{align*}
    \vec{c} := \begin{pmatrix}
        1 - 2 C_{\text{exp, +}}^{u}\\
        -1 + 2 C_{\text{exp, +}}^{\ell}\\
        1 - 2 C_{\text{exp, -}}^{u}\\
        -1 + 2 C_{\text{exp, -}}^{\ell}\\
    \end{pmatrix}
\end{align*}

and

\begin{align*}
    \vec{C} := \begin{pmatrix}
        S_+\\
        -S_+\\
        S_-\\
        -S_-\\
    \end{pmatrix}
\end{align*}

Including both symmetries, the optimization reads

\begin{equation}
    \begin{aligned}
    \mathrm{minimise  }&~~ \frac{1}{d}\sum_{m,n=0}^{d-1} | \bra{mm}\rho\ket{nn}|\\
    \mathrm{s.t.: }&\\
    & \Tr{\rho C(q)}  -  c(q)  \geq 0 ~~~\forall q \in \{1,...,4\} \\
    &\Tr{\rho} -1 = 0 \\
    & \mathcal{T}_n(\rho) = 0 ~~~ \forall n \in \{1,..., d-1\} \\
    & \mathcal{Q}^1_{i}(\rho) = 0 ~~~ \forall i \in \{1,..., d-1\} \\
    & \mathcal{Q}^2_{i,j}(\rho) = 0 ~~~ \forall i \ne j \in \{0,...,d-1\} \\
    & \rho \geq 0 
\end{aligned}
\end{equation}

Note that the objective function contains the sum of absolute values of certain density matrix elements, which is non-linear. Thus, next we rewrite this problem into an equivalent problem with a linear objective function. Therefore, we introduce a family of auxiliary variables $t_{mn}$ for $m,n \in\{0,...,d-1\}$. In order to link $t_{m,n}$ to $|\langle mm|\rho|nn\rangle|$, we further introduce a positive semidefinite matrix $G_{mn}$, which we add as an additional constraint.
This leads to the following modified optimization problem

\begin{equation}\label{eq:PrimalProblem}
    \begin{aligned}
    &\mathrm{minimise  } ~~ \frac{1}{d} \sum_{m,n=0}^{d-1} t_{mn}\\
    \mathrm{s.t.: }\\
    & G_{mn} = \begin{pmatrix}
        t_{mn} & \bra{mm}\rho\ket{nn}\\
        \bra{nn}\rho\ket{mm} & t_{mn}
    \end{pmatrix} \geq 0 \\
    &~~~\forall m,n \in \{0,...,d-1\}  :Y_{mn} \in \mathrm{Pos}(2,2)\\
    & \Tr{\rho C(q)}  -  c(q)  \geq 0 ~~~\forall q \in \{1,...,4\} :z_q \geq 0~~~~~~\\
    &\Tr{\rho} -1 = 0 :y_0 \in \mathbb{R}~~~~~~\\
    & \mathcal{T}_n(\rho) = 0 ~~~ \forall n \in \{1,..., d-1\} :\Lambda_n \in \textrm{Sym}(d+2,d+2)\\
    & \mathcal{Q}^1_{i}(\rho) = 0 ~~~ \forall i \in \{1,..., d-1\} :a_i \in \mathbb{R}~~~~~~\\
    & \mathcal{Q}^2_{i,j}(\rho) = 0 ~~~ \forall i,j \in \{0,...,d-1\} :b_{i,j} \in \mathbb{R}~~~~\\
    & \rho \geq 0 :Z_0 \geq 0~~~~~
\end{aligned}
\end{equation}

where we associated suitable dual variables to each of the constraints, indicated by the variables after the double dot. Finally, we use the method of Lagrange multipliers to obtain the following dual semidefinite program, which provides a certified lower bound on the primal

\begin{equation}\label{eq:DualProblem}
 \begin{aligned}
    &\mathrm{maximise  }~~ \sum_{q=1}^{4} z_q c(q) + y_0 \\
    &\mathrm{s.t.: }&\\
    & \Tr{Y_{mn}} = \frac{1}{d} ~~~\forall m,n \in \{0,...,d-1\}\\
    &- \sum_{m,n=0}^{d-1} K_{mn}(Y_{mn}) - \sum_{q=1}^{4} z_q C(q) - y_0 \mathbbm{1} + \sum_{i=1}^{d-1} a_i (Q(i,i)-Q(0,0)) + \sum_{i,j=0}^{d-1} b_{i,j} (Q(i,j) - Q(0,1)) + \sum_n \mathcal{T}_n^{\dagger}(\Lambda_n)\geq 0\\
    &z_q \geq 0~~~ \forall q \in \{1,...,4\}\\
    &y_0 \in \mathbb{R}\\
    &a_i \in \mathbb{R} ~~~\forall i \in \{1,...,d-1\}\\
    &b_{i,j} \in \mathbbm{R}~~~ \forall i,j \in \{0,...,d-1\}\\
    &\Lambda_n \in \mathrm{Sym}(d+2,d+2),\\
    & Y_{m,n} \geq 0 ~~~ \forall m,n\in \{0,...,d-1\}
\end{aligned}   
\end{equation}

where the dual map of $\mathcal{T}_n$ is given by $\mathcal{T}_n^{\dagger}(X) = T_n X T_n^{\dagger} - T_0 X T_0^{\dagger} $ and we introduced the short notation $K_{m,n}(X) = \ket{ii}\!\!\bra{jj} \bra{ii}X\ket{jj} + \ket{jj}\!\!\bra{ii} \bra{jj}X\ket{ii}$. We model this dual SDP using YALMIP \cite{Lofberg2004} and solve it using MOSEK Version 9.1.9 \cite{mosek}, obtaining a reliable lower bound on our witness. As derived in Section \ref{apdx:derivationWitness}, for a state $\rho$ of dimension $d$ the witness value is related to the Schmidt-rank $k$ of $\rho$ via the inequality $W(\rho) \leq \frac{k}{d}$. This relation allows us to interpret the computed lower bound as a certificate for a minimum Schmidt number and thus a lower bound on the entanglement dimensionality of the source state, $W(\rho) d \leq k.$

\section{Maximal entanglement assumption}\label{apdx:maxent}
Here, we consider the case where the state is maximally entangled in the form of Eq.~\eqref{targetstate} (coming from a purity assumption and energy conservation) \cite{de2002creating}. Even in this case, the boundary effects from our transformation limit the attainable observed visibility in our measurement. By definition, the fidelity of the state (with a maximally entangled state) will be 1, and the only parameter to retrieve is $d$. Based on this pure state assumption and the evolutions leading to Eq.~\eqref{finalentstate}, one can straightforwardly recover a relation between the dimension $d$ and the peak (or dip) visibility as $V=(d-1)/d$, without any need to resort to an SDP. Averaging the visibility of both the peak and dip measurement, we have $V^{\mathrm{exp}}_{\mathrm{avg}} = 0.990\pm0.002$. Using this visibility, along with its relation to the state dimension, we can verify a dimension $d^{\mathrm{exp}} = 100^{+25}_{-17}$. Note that this calculation takes the opposite direction from our minimal assumptions method, since we consider the strongest possible assumption on the state, and we do not restrict the analysis to any subspace of the evolution.

\section{Certified entanglement dimension}\label{apdx:CertifiedEntanglementDimension}
\begin{figure}
    \includegraphics[width=\textwidth]{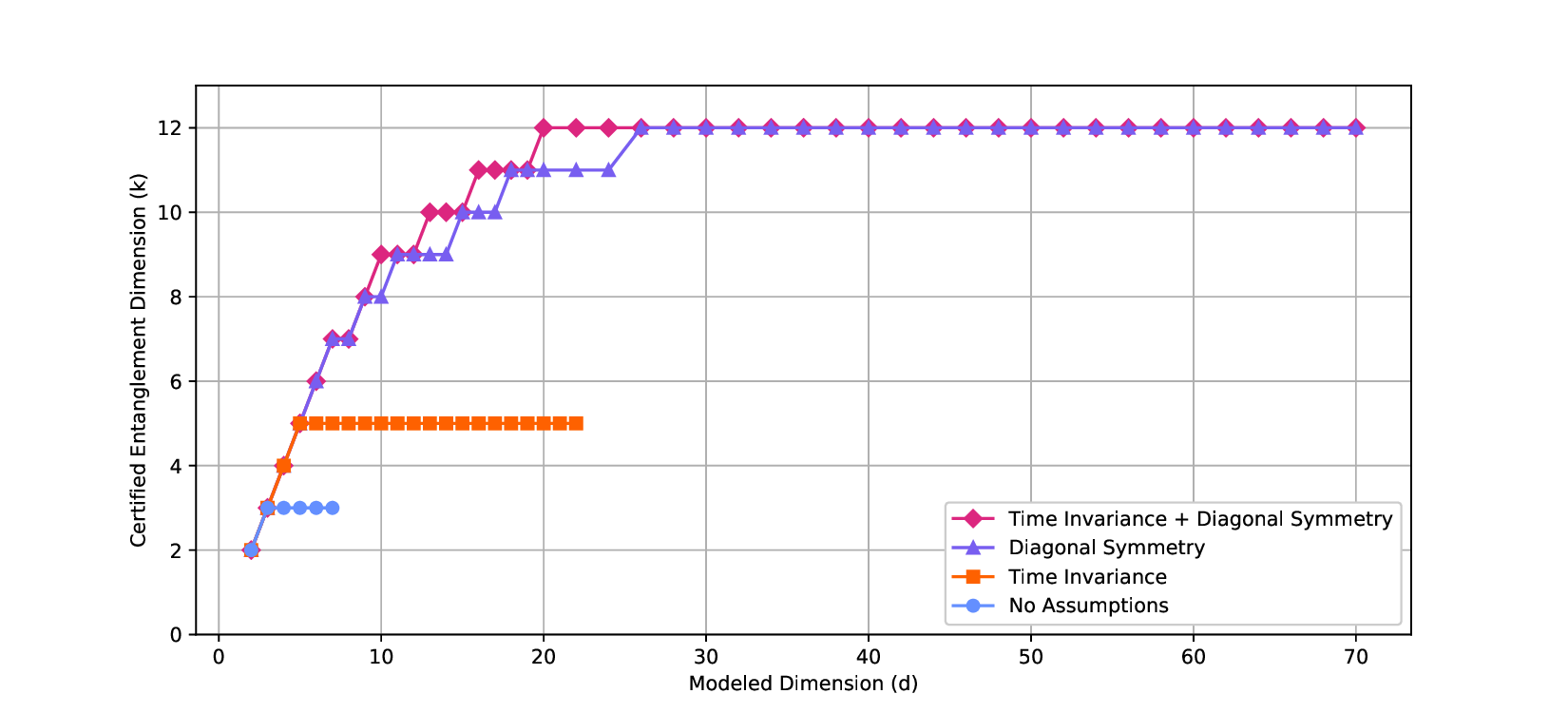}
    \centering
    \caption{Certified entanglement dimension, $k$, as a function of the modeled dimension, $d$ for varying assumptions on the state}
    \label{fig:CertifiedDim}
\end{figure}

Using the SDP above, the certified entanglement dimension, $k$, is evaluated as a function of the matrix dimension, $d$. As illustrated in Fig.~\ref{fig:CertifiedDim}, even in the case with no assumptions, the minimum certified entanglement dimension is already 3, highlighting the presence of high-dimensional entanglement without presuming anything about the state. As physically realistic constraints on the density matrix ($\rho$) are included, the certified dimension also increases.

In addition, by redefining the time bin size during the SDP-based analysis of the data, high-dimensional entanglement can still be certified as the time bins themselves are reduced in size. Table~\ref{tab:timebinreduction} contains the certified entangled dimension for various time bin size reductions. For this certification, both the time-invariance and diagonal symmetry properties were considered for the unknown state. To achieve this reduction in the overall size of the time bins without changing the dimension of the state, the transforms of Eq.~\eqref{symtransforms} are altered so that the number of bins separating each time bin is increased. For example, for the case where the time bins have been reduced by a factor of 4, the new transform of Eq.~\eqref{symtransforms} becomes

\begin{equation}\label{eq:symtransformsreduced}
\begin{split}
    U_\pm\ket{i}_1\ket{j}_2 = \frac{\ket{i}_1\pm\ket{i+8}_1}{\sqrt{2}} \ket{j+4}_2 \;.
\end{split}
\end{equation}

\begin{table}
\caption{Entanglement dimension certification for varied time bin sizes. }
\label{tab:timebinreduction}
\begin{tabular}{lll}

\toprule
Time Bin & Modeled  & Certified Entanglement  \\
Reduction Factor &Dimension (d) &  Dimension (k) \\
\midrule
1 & 20 & 12 \\
2 & 20 & 7 \\
3 & 20 & 6 \\
4 & 20 & 5 \\
\bottomrule

\end{tabular}
\end{table}

\end{document}